\documentclass[twocolumn]{jpsj3}
\usepackage{amsthm,bm,times,mathptmx}

\title{
 Mechanism for the Singlet to Triplet Superconductivity Crossover
\\
 in Quasi-One-Dimensional Organic Conductors
}
\author{Kazuto \textsc{Kajiwara}, 
Masahisa \textsc{Tsuchiizu}, 
Yoshikazu \textsc{Suzumura}, 
and Claude \textsc{Bourbonnais}$^{1}$}

\def\runauthor{K. \textsc{Kajiwara} \it{et al.}}

\inst{Department of Physics, Nagoya University, Nagoya 464-8602\\
$^{1}$D\'epartment de Physique, Universit\'e de Sherbrooke, 
Sherbrooke, Qu\'ebec J1K-2R1, Canada}

\recdate{June 11, 2009; published September 25, 2009 in J.\ Phys.\ Soc.\
Jpn.\ \textbf{78} (2009) 104702}

\abst{ Superconductivity of quasi-one-dimensional organic conductors
 with a quarter-filled band is investigated using the two-loop
 renormalization group approach to the extended Hubbard model for which
 both the single electron hopping $t_{\perp}$ and the repulsive
 interaction $V_{\perp}$ perpendicular to the chains are included.  For
 a four-patches Fermi surface with deviations to perfect nesting, we
 calculate the response functions for the dominant fluctuations and
 possible superconducting states.  By increasing $V_{\perp}$, it is
 shown that a $d$-wave (singlet) to $f$-wave (triplet) superconducting
 state crossover occurs, and is followed by a vanishing spin gap.
 Furthermore, we study the influence of a magnetic field through the
 Zeeman coupling, from which a triplet superconducting state is found to
 emerge.  }

\kword{singlet superconductivity, triplet superconductivity,
organic conductors, extended-Hubbard model, 
 renormalization group, quarter-filled, nesting deviations, Zeeman field}

\begin{document}
\maketitle

\section{Introduction}
Superconductivity in quasi-one-dimensional (quasi-1D) 
organic conductor, (TMTSF)$_2X$, 
has been studied extensively  in the conditions   where   charge  and
spin fluctuations play an important role due to the low dimensionality
of the Fermi surface. \cite{Gruner, Kuroki} 
 The possibility for    triplet state superconductivity, besides the
 singlet one,   is  an    issue of current interest in these materials. 
\cite{Lee,Zhang,Shinagawa1,Yonezawa1}
 From   recent NMR measurements on (TMTSF)$_2$ClO$_4$, it has been
 suggested that spin-triplet  superconductivity may emerge out of a
 singlet state  under   magnetic field. This can occur  as a
 field-induced phase transition that can compete with a FFLO state
 under  strong magnetic field. \cite{Shinagawa1} 
It is  important  for such a study to take into account    the
influence of  low-dimensional fluctuations due to  the strongly
anisotropic band structure of (TMTSF)$_2X$.   The existence
of spin fluctuations  is supported by the NMR experiments  in the normal
phase of these materials.\cite{Wzietek93,Shinagawa1}

 Several theoretical works have been devoted to
  the field-induced phase transition to the triplet 
 superconducting (SC) state. 
Shimahara\cite{Schimahara2000JPSJ,Schimahara2000PRB}
 pointed out such a transition
 using 
 the pairing interactions mediated by spin fluctuations, which 
  have both components for the spin-singlet and triplet pairings.
 Combining   mean-field and RPA methods,
Belmechri {\it et al.} \cite{Belmechri1,Belmechri2}
have shown  that  under field a singlet-triplet superconducting
  transition  is possible, along with the occurrence of a FFLO state at
  intermediate strength of the magnetic field.  
A similar transition has been  shown to occur   by Aizawa {\it et al.}
  \cite{Aizawa1,Aizawa2}, using the RPA method for the extended Hubbard
  model that includes  intersite repulsive interaction  for
  longitudinal and  transverse 
   directions 
 along the chains,
and the Zeeman coupling of spins to a finite magnetic field. 
 As shown for the Hubbard model with on-site repulsive interaction,
 the RPA  approach,
\cite{Scalapino1987,Schimahara1989,Tanaka2004}
which sums up a higher order of perturbation for
 electron interactions, 
 suggests the importance of  the pairing interactions mediated 
   by  spin fluctuations. 
Thus it is  of interest to further examine
 fluctuations of  both  density waves and
  superconducting  pairings, where  
   the  spin gap is essential to the existence 
of a singlet superconducting state.

 These features can be properly taken into account by the   renormalization group (RG) method\cite{Solyom ,Bourbonnais1}.
The effect of  a magnetic field on low-dimensional systems 
 has been studied by the RG method mainly for the one-dimensional cases.
\cite{Penc,Montambaux} 
Noticeable progress has been achieved in studying superconductivity in
zero field for  the case of quasi-1D systems with many chains.
\cite{D-B,Fuseya1,Nickel1}
 However,  the different mechanisms by which  triplet superconductivity can be stabilized in such systems, especially in finite magnetic field,  
have not been fully investigated     within the RG scheme.

In the present work,  we use the RG method  up to the two-loop level
\cite{Tsuchiizu1} 
 to study the competition between 
 the $d$-wave singlet SC  state  (SC$d$) and 
   the $f$-wave triplet SC state (SC$f$) in  quasi-1D 
  systems with  interchain electron hopping and  repulsive interactions.
It is demonstrated that   superconductivity is  driven  by 
 the interplay of  interchain interaction and the nesting deviations.
 The crossover from the SC$d$ state to the SC$f$ state occurs 
with increasing the interchain repulsive interaction  in the presence of nesting deviations.
The effect that a Zeeman  coupling  to the magnetic  field  can have  on
the stability of the SC$d$ state and  the emergence of a triplet SC$f$
state is also studied in details. 
 In \S 2, we give the formulation of the RG technique for  a many-chains
 quasi-1D system at quarter-filling,   in the presence of magnetic field
 and nesting deviations.  
 Using a four-patches decomposition of  the Fermi surface, 
 we  derive the flow equations for     
 the    SC$d$ and   SC$f$ response functions       
  in the superconductivity channel, and for the spin-density-wave  (SDW)  and 
  charge-density-wave (CDW)  responses  in the staggered density-wave channel.
  In \S 3, the results for the possible states  as a function of
  interchain  Coulomb interaction, nesting  deviations and magnetic
  field  are presented.  The conditions for the stability of the SC$f$
  triplet state in the calculated  phase diagrams are given. 
  Summary and discussion are presented  in \S 4.

\section{Formulation}
\subsection{Model}

\begin{figure}[tb]
\begin{center}
\includegraphics[width=3.8cm]{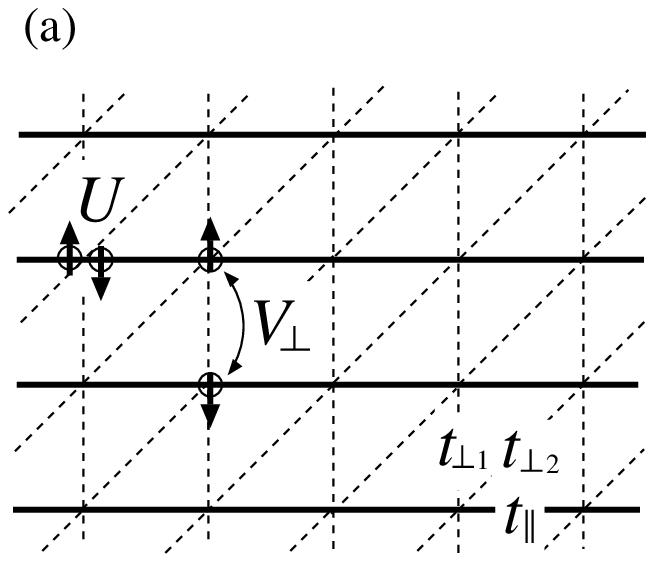}
\includegraphics[width=4.5cm]{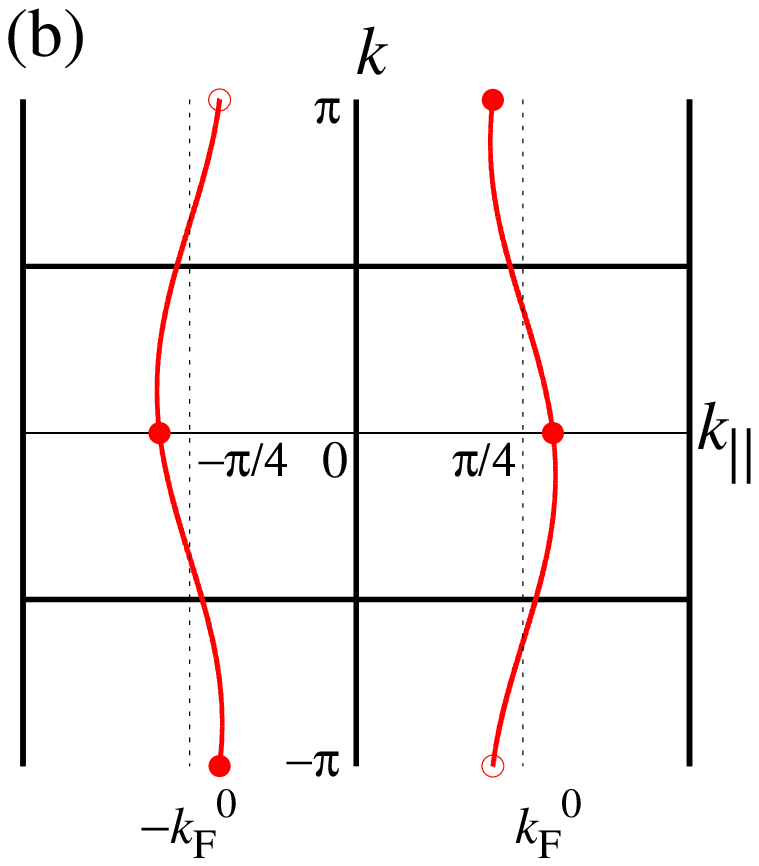}
\end{center}

\vspace*{-.3cm}
\caption{
(Color online)
(a) Extended Hubbard model with intra- and interchain repulsive interactions. 
 Interchain hoppings are shown by dashed lines.   
 (b) The quasi-1D Fermi surface. 
The locations of the patches, 4 in number, are shown by the closed circles.}
\label{fig:model}
\end{figure}

In order to study the superconductivity for the (TMTSF)$_2X$ salt, 
we consider the quasi-1D extended Hubbard model, given by 
\begin{align}
\label{eq:total_H}
{ H}={ H}_0+{ H}_\mathrm{I}, 
\end{align}
where 
\begin{align}
{H}_0
&=
-t_{\|}\sum_{j,l,\sigma}
 (c_{j,l,\sigma}^{\dagger}c_{j+1,l,\sigma}+\mathrm{H.c.})\nonumber\\&
 -t_{\perp1}\sum_{j,l,\sigma}
 (c_{j,l,\sigma}^{\dagger}c_{j,l+1,\sigma}+\mathrm{H.c.}) \nonumber\\&
-t_{\perp 2}\sum_{j,l,\sigma}
 (c_{j,l,\sigma}^{\dagger}c_{j+1,l+1,\sigma}+\mathrm{H.c.})\nonumber\\&
-\mu\sum_{j,l,\sigma}c_{j,l,\sigma}^{\dagger}c_{j,l,\sigma}
 -\frac{h}{2}(n_{j,l,\uparrow}-n_{j,l,\downarrow}),
\label{eq:H0}
\\
{H}_\mathrm{I}
&=
U\sum_{j,l} n_{j,l,\uparrow}n_{j,l,\downarrow}
+V_{\perp}\sum_{j,l}n_{j,l}n_{j,l+1}. 
\label{eq:H_I}
\end{align}
 Here  $c_{j,l,\sigma} (c^{\dagger}_{j,l,\sigma})$ is 
  the  annihilation (creation) operator of an electron 
  at the  site $j$  and chain  $l$, 
     with spin $\sigma$ ($\sigma=\uparrow, \downarrow$), and 
      $n_{j,l,\sigma}=c_{j,l,\sigma}^{\dagger}c_{j,l,\sigma}$.
As illustrated in Fig. \ref{fig:model} (a), 
 the quantities $t_{\|}, t_{\perp1}, t_{\perp2},
  h, \mu, U$ and $V_{\perp}$ are in order  the 
  nearest-neighbor intrachain  and interchain hoppings,
  next-to-nearest-neighbor interchain hopping,   Zeeman field
  ($\mu_\mathrm{B} = 1$),  
  chemical potential, and finally the on-site and interchain repulsive
  interactions.  

Equation (\ref{eq:total_H}) is rewritten 
by making use of the Fourier transform,  
$c_{j,l,\sigma}={(LN_\perp)}^{-1/2}\sum_{\bm k}
 e^{ik_{\|}j+ikl}c_{\sigma}(\bm k)$, where the lattice constant is
taken  as  unity. 
The kinetic energy, eq. (\ref{eq:H0})  can then be  written as  
${H}_0=\sum_{{\bm k},\sigma}{\epsilon}(\bm k)
  c^{\dagger}_{\sigma}(\bm k)c_{\sigma}(\bm k)$. 
Here 
$N_{\perp}$ is the  number of chains, $L$ is their length, 
 ${\bm k}=(k_{\|},k)$ and  
  $\epsilon(\bm k)=
  -2t_{\|}\cos k_{\|} -2t_{\perp1}\cos k -2t_{\perp2}\cos(k_{\|}+k)$.
 In the quasi-1D case,
 we have 
the open Fermi surface since $|t_{\perp1}|,|t_{\perp2}|\ll t_{\|}$.
The Fermi surface, which is a function of $k$, is divided into two parts
for right-going  ($p = +$)  and
 left-going electrons ($p=-$) [see Fig. \ref{fig:model} (b)].   
 Further by adopting the linear-dispersion relation,
 eq. (\ref{eq:H0}) is rewritten as
\begin{align}
{H}_0&=\sum_{{\bm k},\sigma}{\epsilon}_{p,\sigma}(\bm k)c^{\dagger}_{p,\sigma}(\bm k)c_{p,\sigma}(\bm k),\\
{\epsilon}_{p,\sigma}({\bm k})&=v[pk_{\|}-k_{\mathrm{F},\sigma}^{p}(k)],\label{eq:dispersion}\\
k_{\mathrm{F},\sigma}^{p}(k)&=k_\mathrm{F}^0+\sigma \frac{h}{2v}+\frac{2t_{1}}{v}\cos{k}-p\frac{2t_{2}}{v}\sin{k}\label{eq:Fermi-momentum},
\end{align}
where $p=+/-$ (or R/L),  
  $k_\mathrm{F}^{0}=\pi/4$, $v={\sqrt{2}}t_{\|}$,
 $\sigma = +/- (= \uparrow\!/\!\downarrow)$, 
     $t_{1}\equiv t_{\perp1}+t_{\perp2}/\sqrt{2}$, and  
     $t_{2}\equiv t_{\perp2}/\sqrt{2}$.
 Terms of $O(t_{\perp}^2)$ and $k $ dependence 
  of the Fermi velocity are   discarded.

Expressing  the interaction in terms of  
   forward  and  backward scattering \cite{Solyom}, 
  eq. (\ref{eq:H_I}) is rewritten as 
\begin{align}
&
H_\mathrm{I}
=\frac{2\pi v}{LN_\perp}
\sum_{\bm{k_{1},k_{2},q},\{ \sigma \}}
G_{\{\sigma\}(q ,k_1 ,k_2)}
\nonumber\\
&\times 
c_{R,\sigma_1}^{\dagger}(\bm{k_1})c_{L,\sigma_2}(\bm{k_1-q})
c_{L,\sigma_3}^{\dagger}(\bm{k_2-q})c_{R,\sigma_4}(\bm{k_2}) ,
\end{align}
where
\begin{align}
G_{\{\sigma\}(q,k_{1},k_{2})}&=G_{1\perp(q,k_1,k_{ 2})}\delta_{\sigma_1,\sigma_2}\delta_{\sigma_3,\sigma_4}\delta_{\sigma_1,-\sigma_3}\nonumber\\
&-G_{2\perp(q,k_1,k_{ 2})}\delta_{\sigma_1,\sigma_4}\delta_{\sigma_2,\sigma_3}\delta_{\sigma_1,-\sigma_3}\nonumber\\
&-G_{\|(q,k_1,k_{ 2})}\delta_{\sigma_1,\sigma_4}\delta_{\sigma_2,\sigma_3}\delta_{\sigma_1,\sigma_3} .
\end{align}
The couplings $G_{2\perp}$ and $G_{\|}$   stand  for
 the  forward scattering with spin being 
anti-parallel
  and 
 parallel, respectively. 
 The  amplitude  $G_{1\perp}$ denotes that of the  backward scattering 
with anti-parallel spins. 
The 8$k_\mathrm{F}$-Umklapp scattering 
due to the quarter-filling  is  discarded. 
  The coupling constants depend on the wave vector perpendicular to the chains,
\cite{D-B}
 and the definition is  the  same as in ref. \citen{Tsuchiizu1}.
The  bare   scattering amplitudes, which  
 correspond to  the initial values for the RG equations,
 are given by 
\begin{subequations}
\begin{align}
\label{eq:coupling_initial_1}
G_{1\perp(q_{},k_{1},k_{2})}&
=\frac{1}{2\pi v}(U+2V_{\perp}^\mathrm{b}\cos{q_{}}),\\
\label{eq:coupling_initial_2}
G_{2\perp(q_{},k_{1},k_{2})}&=\frac{1}{2\pi v}\bigl[U+2V_{\perp}^\mathrm{f} \cos(k_{1}-k_{2})\bigr],\\
G_{\|(q_{},k_{1},k_{2})}&=\frac{1}{2\pi v}\bigl[2V_{\perp}^\mathrm{f} \cos(k_{1}-k_{2})-2V_{\perp}^\mathrm{b}\cos{q_{}}\bigr].
\label{eq:coupling_initial}
\end{align}
\end{subequations}
In  the following, we shall only retain the  backscattering part
($V_{\perp}^\mathrm{b}$) and neglect the forward scattering contribution
$V_{\perp}^\mathrm{f}$    of
eqs. (\ref{eq:coupling_initial_1})-(\ref{eq:coupling_initial});
$V_{\perp}^\mathrm{b}$  is known to be involved in the enhancement of
$2k_\mathrm{F}^0$ CDW fluctuations.  \cite{Nickel1,Abramovici1}

\subsection{RG equations for the vertex couplings}

\begin{figure}[t]
\begin{center}
\includegraphics[width=6cm]{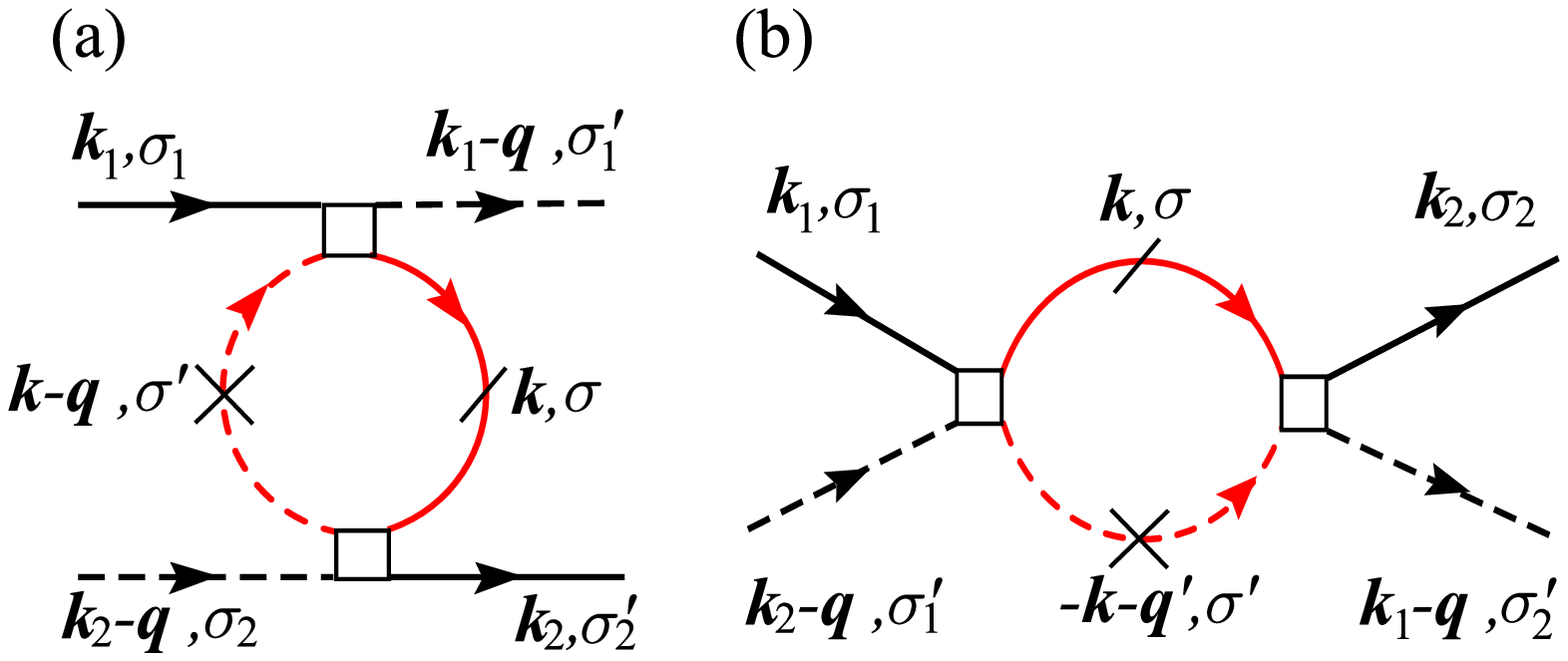}
\end{center}

\vspace*{-.3cm}
\caption{
(Color online)
Peierls (a) and  Cooper (b) bubbles. 
Solid (broken) line represent right (left) going electron, $p=+(-)$; the
 square is the interaction vertex i.e., $G_{1\perp}$, $G_{2\perp}$,
and
 $G_{\|}$. Slashed lines represent  electrons  on high-energy shell
 $E_{l+dl}<|\epsilon_{p,\sigma}(\bm k)|<E_l$, while  crossed lines
 represent electrons at higher energy or lower $l$ values. 
}
\label{fig:bubble}
\end{figure}

We consider the partition function, which is represented in  
 the path integral form, 
\begin{align}
Z=\int\int\mathcal D\psi^{*}\mathcal D\psi \ e^{S[\psi^*,\psi]}
\label{eq:partition-func},
\end{align}
where 
 $S[\psi^*,\psi]=S_0[\psi^*,\psi]+S_\mathrm{I}[\psi^*,\psi]$
 is the action   corresponding   to  the Hamiltonian (\ref{eq:total_H}). 
The   fields  $\psi^{(*)}$ are  the Grassmann variables for the 
 electron degrees of freedom and
$D\psi^{*}\mathcal D\psi$ corresponds to
 the integration measure for the Grassmann variables. 
 In the Fourier-Matsubara space, the free and interacting parts of the action $S_0[\psi^*,\psi]$ and  $S_\mathrm{I}[\psi^*,\psi]$ are respectively given by 
\cite{Bourbonnais1} 
 \begin{align}
S_0[\psi^*,\psi]&={\sum_{{{\bm k},i\omega_n},\sigma,p}\bigl[
  g_{p,\sigma}^0({{\bm
  k},i\omega_n})\bigr]^{-1}\psi^{*}_{p,\sigma}({\tilde{\bm
  k}})\psi_{p,\sigma}({\tilde{\bm k}})},
\\
S_\mathrm{I}[\psi^*,\psi]&
=
-T
\sum_{i\omega_{n_1},i\omega_{n_2},i\omega_m}
 H_\mathrm{I}^\mathrm{G}[{\psi^*,\psi}],
\end{align}
where $T$ is the temperature ($k_B=1$ throughout this work)
  and $H_\mathrm{I}^\mathrm{G}[{\psi^*,\psi}]$ is obtained by substituting the
  fermion operators for the Grassmann variable e.g.,
  $c_{p,\sigma}(\bm{k-q}) \rightarrow
  \psi_{p,\sigma}(\tilde{\bm{k}}-\tilde{\bm{q}})$, etc., where 
${\tilde{\bm k}}=(k_{\|},k_{},\omega_n)$ and
${\tilde{\bm q}}=(q_{\|},q_{},\omega_m)$, with
$\omega_n$ and $\omega_m$ being the Matsubara frequencies for 
 fermions and bosons, respectively. 
The Green's function for the free fermions is given by 
$g_{p,\sigma}^0({\bm k},i\omega_n)
 =\bigl[i\omega_n -{\epsilon_{p,\sigma}(\bm k)}\bigr]^{-1}$. 
In order to examine the behavior at low temperature, 
  we  proceed to implement the successive partial integrations of  
 eq. (\ref{eq:partition-func})  on 
     high-energy shells. This leads to the renormalization of the inner
  shell action  ($S_<$) at step $l$ of the procedure, where $l\ge 0$ is
  a RG parameter for the scaled energy $E_l= E e^{-l}$ ($2E_l$ is the
  scaled bandwidth at $l$).\cite{D-B}  The partial integration of
  high-energy outer shell degrees of freedom is performed perturbatively
  with respect to $S_0$. The use of the linked cluster expansion allows
  one to write the result in the form  
\begin{align}
Z\propto 
\int\int_{<}{\cal D}\psi^*{\cal D}\psi 
\exp\left[
S_{<}[\psi^*,\psi]
+
\sum_{n=1}^{\infty}\frac{1}{n!}\langle S_{\mathrm{I}>}^n\rangle_{0,\mathrm{c}}
\right],
\end{align}
where $\langle \cdots \rangle_{0,\mathrm{c}}$ denotes the  contribution of
connected diagrams  to the outer energy  shell of width $E_ldl$, where $
dl \ll 1$. The action $S_{l+dl}$, at the step $l+dl$, thus contains
additional renormalization  terms, which for the coupling constants read  
\begin{align}
\delta G_{\nu(q,k_1,k_2)}(l)&=G_{\nu(q,k_1,k_2)}(l+dl)-G_{\nu(q,k_1,k_2)}(l) 
\nonumber \\ &
 =\delta G^{n=2}_{\nu(q,k_1,k_2)}+\delta G^{n=3}_{\nu(q,k_1,k_2)}+\cdots ,
\end{align}
where $\nu=1\perp, 2\perp$, and $\|$. 
 The  logarithmic contributions at the one-loop level ($n=2$)  
for  $\delta G^{n=2}_{\nu(q,k_1,k_2)}$ are  given  by the diagrams shown in Fig. \ref{fig:bubble}.

This renormalization consists of two parts, which  come from the 
  Peierls channel and Cooper channels, namely 
\begin{align}
\delta G^{n=2}_{\nu(q,k_1,k_2)}
=
F^{\nu}_{\mathrm{P}(q,k_1,k_2)}+F^{\nu}_{\mathrm{C}(q,k_1,k_2)}.
\end{align}
The  Peierls bubble $F^{\nu}_\mathrm{P}(q,k_1,k_2)$ is given by
\begin{align}
&F^{\nu}_{\mathrm{P}(q,k_1,k_2)}= -\frac{2 \pi vT}{LN_\perp}
 \sum_{\nu_1,\nu_2}
{\sum^{\mathrm{o.s.}}_{\bm k}}'
\sum_{i\omega_n} \nonumber\\
&\quad \times
J^{\nu,\nu_1,\nu_2}_{\mathrm{P}(q,k,k_1,k_2)}g^0_{R,\sigma}(\bm
 k,i\omega_n)g^0_{L,\sigma'}({\bm{ k-q}},i\omega_n),
\label{eq:def-Peierls-bubble}
\end{align}
where 
\begin{align}
J^{\nu,\nu_1,\nu_2}_{\mathrm{P}(q,k,k_1,k_2)}
= 
(-1)^{m_p^{\nu, \nu_1, \nu_2}}G_{\nu_1(q,k_1,k)}G_{\nu_2(q,k,k_2)} ,
\end{align}
 and  the summation of $\nu_1$, $\nu_2$  $( =1\perp, 2\perp, \|$)  is
 taken  for the fixed $\nu$, as shown explicitly later.   
The quantity $m_p^{\nu, \nu_1, \nu_2}$ denotes 
 the number of the permutation of Grassmann variable.
 ${\sum^{\mathrm{o.s.}}_{\bm k}}'$ is performed  in the outer shell region.
The summation is written as   
\begin{align}
{\sum^{o.s.}_{\bm k}}'=2\sum^{o.s.}_{\bm k}\Theta (|\epsilon_{L,\sigma'}(\bm{k-q})|-E_l)\label{eq:summation-P},
\end{align}
where $\sum^{o.s.}_{\bm k}$ denotes the summation in the region of 
$E_{l+dl}<|\epsilon_{R,\sigma}(\bm k)|<E_l$
and $\Theta (x) $  is the Heaviside step function 
with the definition $\Theta (0)\equiv 1/2$. 
 In the RG procedure, the external momentum is fixed at the Fermi 
surface, i.e., 
$(k_{\mathrm{F},\sigma_1}^R(k_{1}),k_{1})$, 
$(-k^L_{\mathrm{F},\sigma_2'}(k_{2}-q_{}),k_{2}-q_{})$ 
for the incoming states and 
$(k^R_{\mathrm{F},\sigma_2}(k_{2}),k_{2})$, 
$(-k^L_{\mathrm{F},\sigma_1'}(k_{1}-q_{}),k_{1}-q_{})$
 for the outgoing states. 
Thus  for the momentum summation of the bubble with 
  ${\bm q} = (q_{\|}, q)$, 
 the longitudinal momentum $q_{\|}$ is determined by 
  the momentum conservation for the respective vertex ($\nu_1, \nu_2$).
\cite{Tsuchiizu1} 
Using  $(LN_\perp)^{-1}\sum^{\mathrm{o.s.}}_{\bm k}={(2\pi v N_{\perp})}^{-1}\sum_{k}\int_{o.s.}d\epsilon_{R,\sigma}(\bm k)$, 
 and performing the Matsubara-frequency summation, 
eq. (\ref{eq:def-Peierls-bubble}) is rewritten as 
\begin{align}
&F^{\nu}_{\mathrm{P}(q_{},k_1,k_2)}
=-\frac{2}{N_{\perp}}{\sum_{\nu_1,\nu_2}\sum_{k}}
J^{\nu,\nu_1,\nu_2}_{\mathrm{P}(q,k,k_1,k_2)}\nonumber\\
&
\qquad
\times \int_{o.s.}d\epsilon_{R,\sigma}(\bm k)
 \frac{f(\epsilon_{R,\sigma}(\bm k))-f(\epsilon_{L,\sigma'}({\bm
 {k-q}}))}{\epsilon_{R,\sigma}(\bm k)-\epsilon_{L,\sigma'}({\bm {k-q}})}
\nonumber\\
&
\qquad
\times\Theta(|\epsilon_{L,\sigma'}({\bm{k-q}})|-E_l) , 
\label{eq:Peierls-bubble}
\end{align}
where 
$\int_{o.s.}d\epsilon=\int_{E_{l+dl}}^{E_l}d\epsilon
 +\int_{-E_{l}}^{-E_{l+dl}}d\epsilon$.
 The Fermi distribution function $f(x)$ in eq. (\ref{eq:Peierls-bubble}) 
  is given by   $f(x) = 1/ [\exp (x/ T) + 1]$,  
which will be   treated in the low-temperature limit. 
Following the RG procedure 
 one can rescale  the energy 
 2$E_{l}$ up to the original   band width $2E$, so that 
all the  energies $ A(\equiv  t_{1}, t_{2}, h)$ 
  are rescaled according to  
\begin{align}
 A(l) = A \mathrm{e}^l, 
\end{align}
 where the  anomalous corrections to  $A$ due to 
  the two-loop corrections
 are    small and have been neglected 
  for the present choice of parameters  
(see Appendix). 
At zero temperature, 
eq. (\ref{eq:Peierls-bubble}) is calculated as 
\begin{align}
F^{\nu}_{\mathrm{P}(q_{},k_1,k_2)}
=&
\frac{1}{N_{\perp}}\sum_{\nu_1,\nu_2}\sum_{k} 
 J^{\nu,\nu_1,\nu_2}_{\mathrm{P}(q,k,k_1,k_2)}
\nonumber\\ &\times 
\frac{1}{2}\sum_{i=1,2}
 I_{\mathrm{P}(q_{},k_{},k_{i},\sigma-\sigma_i)}dl ,
\end{align}
where 
\begin{align}
I_{\mathrm{P}(q_{},k_{},k_{i},\sigma-\sigma_i)}
=\frac{2E}{2E+ |Y^\mathrm{P}_{q_{},k_{},k_{ i},\sigma-\sigma_i}(l)|},
\end{align}
and 
\begin{align}
Y^\mathrm{P}_{q_{},k_{},k_{i},\sigma-\sigma_i} (l)
\equiv& -\epsilon_{R,\sigma}(\bm k)-\epsilon_{L,\sigma'}(\bm{k-q})
\nonumber\\
=&
 +2t_{1}(l)[\cos{k}+\cos(k-q)]
\nonumber\\&
 -2t_{1}(l)[\cos{k_i}+\cos(k_i-q)]
\nonumber\\&
 -2t_{2}(l)[\sin{k_{}-\sin{(k-q)}}]
\nonumber\\&
 +2t_{2}(l)[\sin{k_{i}+\sin{(k_i-q)}}]
\nonumber\\&
 +h(l)(\sigma-\sigma_i)
\label{eq:YP}.
\end{align}
In the above equations, $\sigma = \uparrow$ and $\downarrow$ are 
 replaced by $\sigma = + 1$ and $-1$, respectively.  
In a similar way, the Cooper bubble is calculated as follows:
\begin{align}
&F^{\nu}_{\mathrm{C}(q,k_1,k_2)}
=
-\frac{2\pi vT}{LN_\perp} 
 \sum_{\nu_1,\nu_2}{\sum^{\mathrm{o.s.}}_{\bm k}}''
 \sum_{i\omega_n}\nonumber\\
&\quad
\times
 J^{\nu,\nu_1,\nu_2}_{\mathrm{C}(q,k,k_1,k_2)}
 g^0_{R,\sigma}(\bm k,i\omega_n)g^0_{L,\sigma'}({\bm{-k-q'}},-i\omega_n) , 
\label{eq:def-Cooper-bubble}
\end{align}
where $\bm q'\equiv \bm {q-k_1-k_2}$ and
\begin{align}
J^{\nu,\nu_1,\nu_2}_{\mathrm{C}(q,k,k_1,k_2)}
=
(-1)^{m_p^{\nu,\nu_1,\nu_2}}G_{\nu_1(q-k_2+k,k_1,k)}
G_{\nu_2(q-k_1+k,k,k_2)},
\end{align}
and the summation ${\sum^{\mathrm{o.s.}}_{\bm k}}''$ is given by
substituting $\epsilon_{L,\sigma'}(\bm{-k-q'})$ for
$\epsilon_{L,\sigma'}(\bm {k-q})$ in eq. (\ref{eq:summation-P}). 
The Fermi surface of the external momentum is given by 
$(k^R_{\mathrm{F},\sigma_1}(k_{1}),k_{1})$, $(-k^L_{\mathrm{F},\sigma_1'}(k_{2}-q_{}),k_{2}-q_{})$ for the incoming states, and 
$(k^R_{\mathrm{F},\sigma_2}(k_{2}),k_{2})$,
$(-k^L_{\mathrm{F}, \sigma_2'}(k_{1}-q_{}),k_{1}-q_{})$ 
for the outgoing states. 
The longitudinal momentum vector $q'_{\|}$ is also determined by 
 the  momentum conservation for respective vertex. 
Performing the Matsubara-frequency summation, 
eq. (\ref{eq:def-Cooper-bubble}) is rewritten as
\begin{align}
&F^{\nu}_{\mathrm{C}(q,k_1,k_2)}=
 \frac{2}{N_{\perp}}\sum_{k}
 \sum_{\nu_1,\nu_2}J^{\nu,\nu_1,\nu_2}_{\mathrm{C}(q,k,k_1,k_2)}
\nonumber\\
& \qquad 
\times 
 \int_\mathrm{o.s.}d\epsilon_{R,\sigma}(\bm k)
 \frac{f(\epsilon_{R,\sigma}(\bm k))
          -f(-\epsilon_{L,\sigma'}({\bm {-k-q'}}))}
      {\epsilon_{R,\sigma}(\bm k)+\epsilon_{L,\sigma'}({\bm {-k-q'}})}
\nonumber\\
&\qquad
\times\Theta(|\epsilon_{L,\sigma'}({\bm{-k-q'}})|-E_l).
\label{eq:Cooper-bubble}
\end{align}
At   zero temperature, eq. (\ref{eq:Cooper-bubble}) is calculated as
\begin{align}
F^{\nu}_{\mathrm{C}(q,k_1,k_2)}
=&
-\frac{1}{N_{\perp}}
\sum_{\nu_1,\nu_2}
\sum_{k}J^{\nu,\nu_1,\nu_2}_{\mathrm{C}(q,k,k_1,k_2)}
\nonumber\\
&\times \frac{1}{2}\sum_{i=1,2}{I}_{\mathrm{C}(q',k,k_{i},\sigma-\sigma_i)},
\end{align}
where 
\begin{align}
I_{\mathrm{C}(q',k,k_{i},\sigma-\sigma_i)}
&=\frac{2E}{2E+ |Y^\mathrm{C}_{q'_{},k_{},k_{i},\sigma-\sigma_i}(l)|},
\end{align}
and 
\begin{eqnarray} 
Y^\mathrm{C}_{q'_{},k_{},k_{i},\sigma-\sigma_i} (l)
&\equiv& 
\epsilon_{R,\sigma}(\bm k)+\epsilon_{L,\sigma'}(\bm{-k-q'})
\nonumber\\
&=&
 +2t_{1}(l)[\cos{k}-\cos(k+q')]
\nonumber\\&&
 -2t_{1}(l)[\cos{k_i}-\cos(k_i+q')]
\nonumber\\&&
-2t_{2}(l)[\sin{k_{}-\sin{(k+q')}}]
\nonumber\\&&
 +2t_{2}(l)[\sin{k_{i}-\sin{(k_i+q')}}]
\nonumber\\&&
 +h(l)(\sigma-\sigma_i)
\label{eq:YC}.
\end{eqnarray}
The  functions  
$I_{\mathrm{P}(q,k,k_i,\sigma-\sigma_i)}$ and 
 $I_{\mathrm{C}(q,k,k_i,\sigma-\sigma_i)}$,  which are equal to 
 unity in  the one-dimensional case and  in the absence of the magnetic 
field, 
are reduced by $t_{1}$ and $t_{2}$ 
in the quasi-1D case. 

\begin{figure}[t]
\begin{center}
\includegraphics[width=7cm]{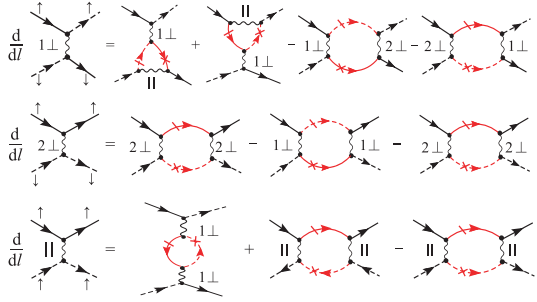}
\end{center}

\vspace*{-.3cm}
\caption{
(Color online)
The RG equations  for the coupling constants at the one-loop level. 
The slashed and crossed internal lines represent the electrons on the 
 high-energy shell and within the shell, respectively.
The diagrams with  the slash and cross  lines  interchanged (not shown) 
  are also taken into account.
}
\label{fig:diagram}
\end{figure}

 The RG flow equations at the one-loop level are  shown diagrammatically  in 
Fig. \ref{fig:diagram}. 
Including the two-loop corrections  derived in the Appendix,
 the  RG equations  take the form 
\begin{align}
\dfrac{{d}}{{dl}}G_{\nu(q,k_1,k_2)}
&=
\frac{1}{2N_{\perp}}\sum_{k}\Xi^{(1)}_{\nu(q,k,k_1,k_2)}
\nonumber\\
&
-\frac{1}{8N^2_{\perp}}G_{\nu}(q,k_1,k_2)
\sum_{q',k'}\Xi^{(2)}_{\Sigma(q,k_1,k_2,q',k')}
\nonumber\\
&
+\frac{1}{4N^2_{\perp}}\sum_{q',k'}
\Xi^{(2)}_{\nu (q,k_1,k_2,q',k')}+(k_1\leftrightarrow k_2),
\label{full:RG}
\end{align}
where $\Xi^{(1)}_{\nu(q,k,k_1,k_2)}$ is 
the contribution from the one-loop RG, while 
 $\Xi^{(2)}_{\Sigma (q,k_1,k_2,q',k')}$ and 
 $\Xi^{(2)}_{\nu (q,k_1,k_2,q',k')}$ are the two-loop  contributions 
 coming from the self-energy and the vertex part,
 respectively (see Appendix).
\cite{Tsuchiizu1} 
The quantity $\Xi^{(1)}_{\nu(q,k,k_1,k_2)}$
 is written   as 
\begin{subequations}
\begin{align}
&\Xi^{(1)}_{1\perp(q,k,k_1,k_2)}
=
G_{1\perp(q,k_1,k)}G_{\|(q,k,k_2)}{I}_{\mathrm{P} (q,k,k_1,k_2)}^{(1)}
\nonumber\\
& \qquad
+G_{\|(q,k_1,k)}G_{1\perp(q,k,k_2)}{I}_{\mathrm{P} (q,k,k_2,k_1)}^{(1)}
\nonumber\\
&  \qquad
-G_{1\perp(q-k_2+k,k_1,k)}G_{2\perp(q-k_1+k,k,k_2)}
  I_{\mathrm{C} (q-k_1-k_2,k,k_1,k_2)}^{(1)}
\nonumber\\
&  \qquad
-G_{2\perp(q-k_2+k,k_1,k)}G_{1\perp(q-k_1+k,k,k_2)}
I_{\mathrm{C} (q-k_1-k_2,k,k_2,k_1)}^{(1)},
\\
&\Xi^{(1)}_{2\perp(q,k,k_1,k_2)}
=
G_{2\perp(q,k_1,k)}G_{2\perp(q,k,k_2)}
I_{\mathrm{P}(q,k,k_1,k_2)}^{(0)}
\nonumber\\
&  \qquad
-G_{1\perp(q-k_2+k,k_1,k)}G_{1\perp(q-k_1+k,k,k_2)}
I_{\mathrm{C}(q-k_1-k_2,k,k_1,k_2)}^{(2)} 
\nonumber\\
&  \qquad
-G_{2\perp(q-k_2+k,k_1,k)}G_{2\perp(q-k_1+k,k,k_2)}
I_{\mathrm{C}(q-k_1-k_2,k,k_1,k_2)}^{(0)},
\\
&
\Xi^{(1)}_{\|(q,k,k_1,k_2)}
=
G_{1\perp(q,k_1,k)}G_{1\perp(q,k,k_2)}
I_{\mathrm{P}(q,k,k_1,k_2)}^{(2)}
\nonumber\\
&  \qquad
+G_{\|(q,k_1,k)}G_{\|(q,k,k_2)}
I_{\mathrm{P}(q,k,k_1,k_2)}^{(0)}
\nonumber\\
&  \qquad
-G_{\|(q-k_2+k,k_1,k)}G_{\|(q-k_1+k,k,k_2)}
I_{\mathrm{C}(q-k_1-k_2,k,k_1,k_2)}^{(0)},
\end{align}
\end{subequations}
where for $\lambda=\mathrm{P}$, $\mathrm{C}$,
\begin{subequations}
\begin{align}
I_{\lambda(q,k,k_1,k_2)}^{(0)}
&=
\frac{1}{2}\sum_{i=1,2}I_{{\lambda}(q,k,k_i,0)},
\\
I_{\lambda(q,k,k_1,k_2)}^{(1)}
&=
\frac{1}{4}\sum_{r=\pm 1}
\big[I_{\lambda({q,k,k_1,2r})}+I_{\lambda(q,k,k_2,0)} \big],
\\
I_{\lambda(q,k,k_1,k_2)}^{(2)}
&=
\frac{1}{4}\sum_{r=\pm1}\sum_{i=1,2}I_{\lambda({q,k,k_i,2r})}.
\end{align}
\end{subequations}
 Note that in the absence 
 of   interchain couplings, $\Xi^{(1)}_{\nu(q,k,k_1,k_2)}$   coincides
 with the expressions already obtained by Montambaux {\it et al.,}
 \cite{Montambaux} at $h\ne 0$ in the 1D case.

\subsection{RG equations for response functions}

Now we calculate the response functions  for CDW, SDW, SC$d$, and SC$f$.
The composite fields of corresponding order parameters  are defined as
\begin{subequations}
\begin{align}
{\cal O}_\mathrm{CDW}({\bm q}_\mathrm{P})
&=
\sqrt{\frac{1}{LN_\perp}}
\sum_{{\bm k},\sigma}
c^{\dagger}_{R,\sigma}(\bm k)c_{L,\sigma}({\bm{k-q}}_\mathrm{P}),
\\
{\cal O}_\mathrm{SDW}({\bm q}_\mathrm{P})
&=
\sqrt{\frac{1}{LN_\perp}}
\sum_{{\bm k},\sigma} 
c^{\dagger}_{R,\sigma}(\bm k)c_{L,-\sigma}({\bm{k-q}}_\mathrm{P}),
\\
{\cal O}_{\mathrm{SC}d}({\bm q}_\mathrm{C})
&=
\sqrt{\frac{1}{LN_\perp}}
\sum_{{\bm k},\sigma}
(\sigma \cos k) \,
 c_{R,\sigma}(\bm k)c_{L,-\sigma}({\bm{-k+q}_\mathrm{C}}),
\label{O-SCd}\\
{\cal O}_{\mathrm{SC}f}({{\bm q}}_\mathrm{C})
&=
\sqrt{\frac{1}{LN_\perp}}
\sum_{{\bm k},\sigma}
(\sin{k_{\|}}\cos{k})\, 
c_{R,\sigma}(\bm k)c_{L,\sigma}({{\bm{-k+q}}_\mathrm{C}}),
\label{O-SCf}
\end{align}
\end{subequations}
where the density-wave modulation 
${\bm q}_\mathrm{P}=(2k_\mathrm{F}^0,\pi)$ 
 is the nesting vector and  ${\bm q}_\mathrm{C}=(0,0)$.

For   simplicity, we use the four patches model  for the Fermi surface, 
  for which   $k_{}=0,\pi$, as shown in Fig. \ref{fig:model} (b).  
We thus end up 
with  12 independent coupling constants. 
 Such a choice   allows a qualitative description of the SC states, i.e.,  
 eqs. (\ref{O-SCd}) and (\ref{O-SCf}) having  respectively the  gap functions  
 $\cos k_{} $ and $\sin k_{\|} \cos k_{} $.  This corresponds to nodes
   located at 
  $k_{}=\pm \pi/2$ for SC$d$,  and $k_{\|}=0$ and $k_{}=\pm \pi/2$
for SC$f$. 
Hereafter, we use  $0$ or $\pi$ for the momentum 
 perpendicular to the chain.
 While the 
 $t_{2}$-terms 
 in eqs. (\ref{eq:YP}) and (\ref{eq:YC})
   vanish for $q,k,k_i =  0, \pi$,
  nesting deviations can be incorporated in the Peierls channel by introducing 
 the following factor for the Peierls channel
 in eq. (\ref{full:RG}), 
 \begin{align}
I_{t_{2}^*}=\frac{E}{E+t_{2}^*(l)},
\label{eq:nesting}
\end{align} 
where   $t_{2}^* (\propto t_{2}$) is a cut-off energy.  It follows that 
 $I_{t_{2}^*}\simeq 1$ for $E \gg t_{2}^*(l)$
 and 
 $I_{t_{2}^*}\simeq 0$ for $E\ll t_{2}^*(l)$. 
A cutoff procedure similar to  eq. (\ref{eq:nesting}) has been used 
 in studying the density-wave problem  in many-coupled-chains case.
\cite{Emery}

We calculate the response functions by adding a linear coupling 
of order parameters  to source fields in the action,\cite{Bourbonnais1}
 that is 
$S_h[\psi,\psi^*,h,h^*]=\sum_{\mu}(h_{\mu}{\cal O}^*_{\mu}+\mathrm{c.c.})$,
 where 
$h_{\mu}$ is a source field in the  
$\mu$=CDW, SDW, SC$d$, and SC$f$ channels.
The total action at step $l$ is given by 
\begin{align}
&S[\psi,\psi^*,h,h^*]_{l}=S[\psi^{*}, \psi ]_l
\nonumber\\
&+\sum_{\mu}(h_{\mu}z_{\mu}(l){\cal O}^*_{\mu}+\mathrm{c.c.})
+\chi_{\mu}(l)h_{\mu}h_{\mu}^*,
\end{align}
where $z_\mu $ is a renormalization factor for the order parameter 
 vertex ($z_{\mu}(0)=1$),
 and  $\chi_\mu(l)$ is the  response function with $\chi_{\mu}(0)=0$.

\begin{figure}[t]
\begin{center}\leavevmode
\includegraphics[width=5.5cm]{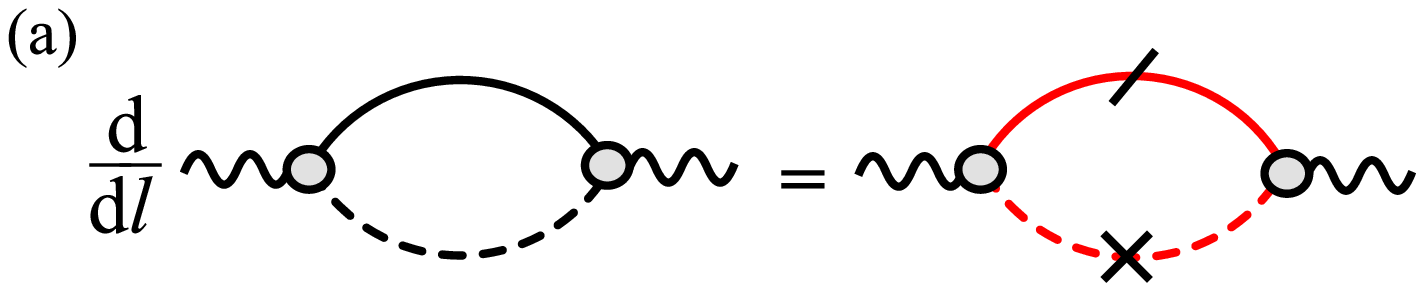}

\vspace*{.5cm}
\includegraphics[width=5.5cm]{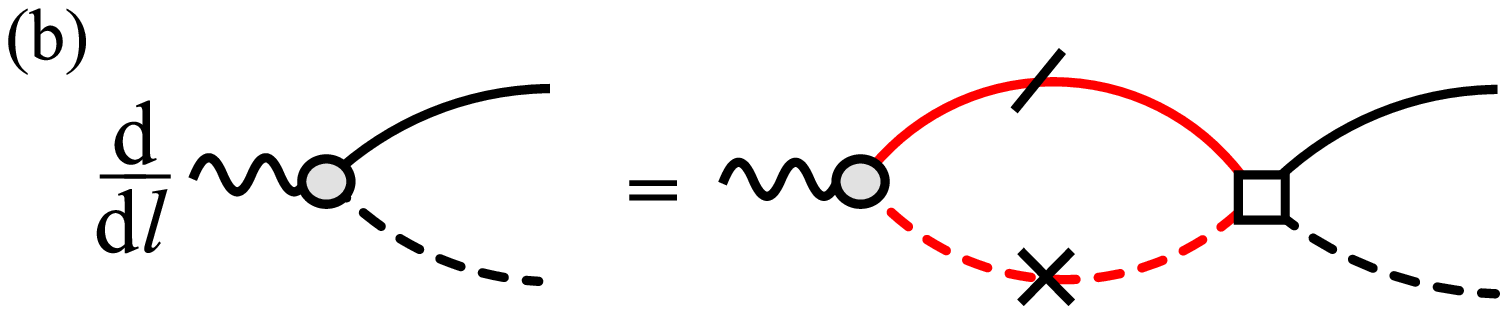}
\end{center}

\vspace*{-.3cm}
\caption{
(Color online)
Renormalization group equations  for the response functions (a), and order parameter vertex part   for the source field (b).}
\label{fig:Feynman-diagram}
\end{figure}

 Up to the one-loop level (see Fig. \ref{fig:Feynman-diagram}), 
the  RG equations 
for these quantities are 
\begin{subequations}
\begin{align}
\frac{d}{dl} \chi_{{\mathrm{CDW}}}({\bm q}_{\mathrm P})
&=\frac{1}{\pi v}z^2_{{\mathrm{CDW}}}({\bm q}_{\mathrm P})I_{h/2}I_{t_{2}^*},
\\
\frac{d}{dl} \chi_{{\mathrm{SDW}}}({\bm q}_{\mathrm P})
&=\frac{1}{\pi v}z^2_{{\mathrm{SDW}}}({\bm q}_{\mathrm P})I_{t_{2}^*},
\\
\frac{d}{dl} \chi_{{\mathrm{SC}}d}({\bm q}_{\mathrm C})
&=\frac{1}{\pi v}z^2_{{\mathrm{SC}}d}({\bm q}_{\mathrm C})I_{h/2},
\\
\frac{d}{dl} \chi_{{\mathrm{SC}}f}({\bm q}_{\mathrm C})
&=\frac{1}{\pi v}z^2_{{\mathrm{SC}}f}({\bm q}_{\mathrm C}),
\end{align}
\end{subequations}
where the respective three-point vertices  $z_\mu$ obey
  the RG equations:
\begin{subequations}
\begin{align}
\frac{d}{dl} {\ln z_{{\mathrm{CDW}}}({\bm q}_{\mathrm P})}
=&\bigl[-G_{1\perp(\pi,0,0)}-G_{1\perp(\pi,0,\pi)}\nonumber\\
&+G_{\|(\pi,0,\pi)}+G_{\|(\pi,0,0)}\bigr]I_{h/2}I_{t_{2}^*},
\label{eq:dlnz-CDW}\\
\frac{d}{dl} {\ln z_{{\mathrm{SDW}}}({\bm q}_{\mathrm P})}
=&\bigl[G_{2\perp(\pi,0,\pi)}+G_{2\perp(\pi,0,0)}\bigr]I_{t_{2}^*},
\label{eq:dlnz-SDW}\\
\frac{d}{dl} {\ln z_{{\mathrm{SC}}d}({\bm q}_{\mathrm C})}
=&[-G_{1\perp(0,0,0)}+G_{1\perp(\pi,0,\pi)}
\nonumber\\
&+G_{2\perp(\pi,0,\pi)}-G_{2\perp(0,0,0)}]I_{h/2}
\label{eq:dlnz-SCd},\\
\frac{d}{dl} {\ln z_{{\mathrm{SC}}f}({\bm q}_{\mathrm C})}
=&G_{\|(\pi,0,\pi)}-G_{\|(0,0,0)} .
\label{eq:dlnz-SCf}
\end{align}
\end{subequations}
The initial values are $z_\mu |_{l=0}=1$.
In the above equations,  $I_h$ is defined by 
\begin{equation}
I_{h}=\frac{E}{E+h(l)} .
\label{eq:Ih}
\end{equation}
Note that  eq. (\ref{eq:dlnz-SDW})
  represents  the flow equation for the transverse SDW. 
The longitudinal SDW response  is obtained by  substituting 
 $[G_{1\perp(\pi,0,0)}+G_{1\perp(\pi,0,\pi)}
 +G_{\|(\pi,0,\pi)}+G_{\|(\pi,0,0)}]I_{h/2}I_{t_{2}^*}$ 
for the r.h.s. of  eq. (\ref{eq:dlnz-SDW}). 
 These two flow equations become equivalent in the absence of $h$.

\section{Singlet versus  triplet superconductivity}

For the numerical calculations that follow, 
parameters of the model are   fixed at  
 $U=4t_{\|}$ and  $t_{1}=0.2t_{\|}$ unless stated explicitly. 
The  unit of the energy is  $t_{\|}$, 
which is  set to unity $t_{\|}=1$, and we take $E$= 2$t_{\|}$. 
The scaling parameter $l$ is replaced by 
  $l=\ln(E/T)$, where $T$ 
can be squared with the actual temperature introduced in \S2.

In the following  study of the four-patches Fermi surface, we will
also examine  the spin gap $\Delta_\sigma$,
  which is governed by the combination of coupling constants
 \cite{Tsuchiizu2005}
\begin{align}
\label{eq:g_sigmaplus}
G_{\sigma+} \equiv 
\frac{1}{2}\bigl[ G_{{2 \perp}(0,0,0)} - G_{{\|} (0,0,0)} 
  + 
 G_{{2 \perp}(\pi,0,0)} - G_{{\|} (\pi,0,0)} \bigr].
\end{align} 
For a non zero  $\Delta_{\sigma}$,
the coupling $G_{\sigma+}$ takes  a positive value at  $l=0$, but 
 moves to a fixed point with a large negative value. 
 At a qualitative level  for the spin gap, we shall use 
$\Delta_{\sigma} = E_{l_{\sigma}} (= E \mathrm{e}^{-l_{\sigma}})$  
 where $l_{\sigma}$ is determined
 by  the condition $G_{\sigma+}(l_{\sigma}) = -0.7$.
\cite{Tsuchiizu1}
Thus   a non zero spin gap is obtained for the couplings, 
 $G_{{2 \perp}(0,0,0)} < 0$,   
 $G_{{\|} (0,0,0)} > 0$,  
 $G_{{2 \perp}(\pi,0,0)} < 0$, and 
 $G_{{\|} (\pi,0,0)} > 0$. 
Here we note that in the absence of magnetic field,  the combination of couplings for the spin gap,
 eq. (\ref{eq:g_sigmaplus}), can be  rewritten as 
\begin{align}
\label{eq:g_sigmaplus0}
G_{\sigma+} = 
 \frac{1}{2}\bigl[ G_{{1 \perp}(0,0,0)} + G_{{1 \perp} (\pi ,0,0)} \bigr].
\end{align}

From $G_{\sigma +}$,  the uniform susceptibility within  RPA 
takes the form \cite{Fuseya2005,Fuseya2007}
\begin{align}
\label{eq:uniform_susceptibilty}
\chi_s =  \frac{\chi_{s0}}{1 - 2\pi vG_{\sigma +}\chi_{s0}} ,
\end{align} 
where $\chi_{s0}$ 
denotes the susceptibility for free electrons given by 
 $2 \pi v \chi_{s0} = \tanh (E/2T)$.

\subsection{Case of perfect nesting ($t_{2}^* = 0)$}

\begin{figure}[t]
\begin{center}\leavevmode
\includegraphics[width=7cm]{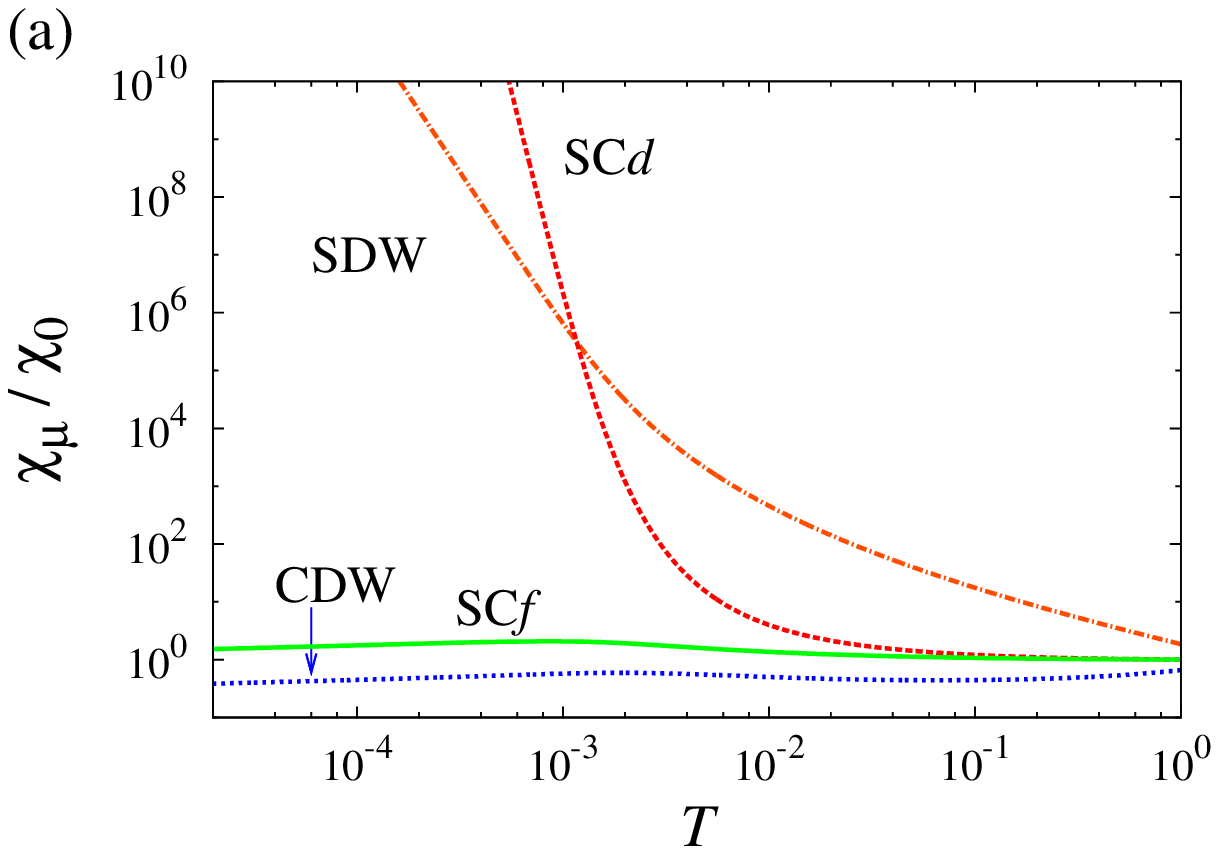}
\includegraphics[width=7cm]{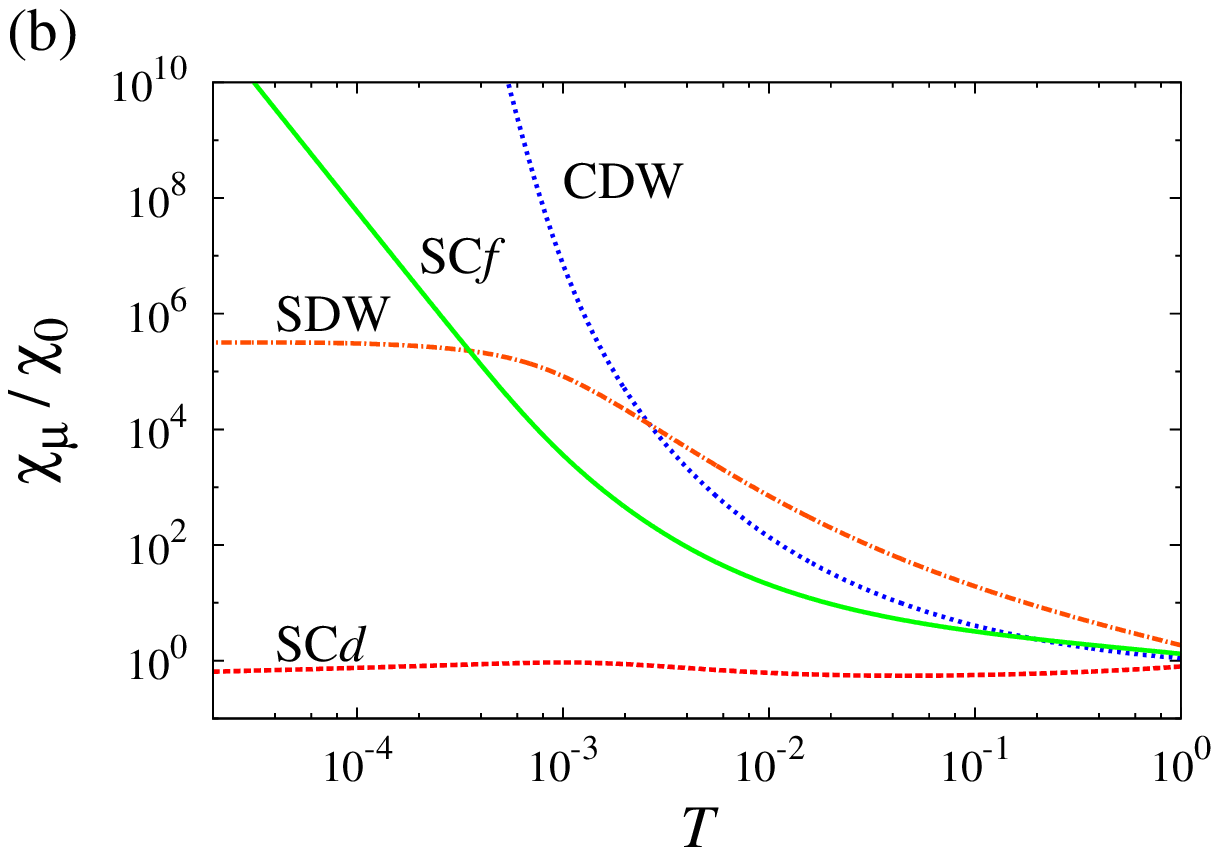}
\end{center}

\vspace*{-.3cm}
\caption{
(Color online)
Temperature dependence of response function with $t_{2}^*=0.0, h=0.0$
 for $V_{\perp}^\mathrm{b}=0.0$ (a), and $1.0$ (b). 
Response functions $\chi_{\mu}$ are
 normalized by noninteracting response function
 $\chi_{\mathrm 0} (=l/\pi v)$.
}
\label{fig:enest0.0-res}
\end{figure}

\begin{figure}[t]
\begin{center}
\includegraphics[width=7cm]{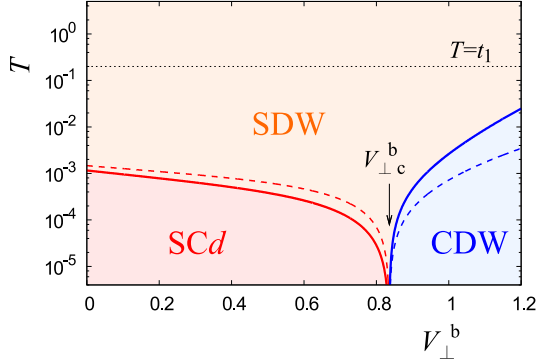}
\end{center}

\vspace*{-.3cm}
\caption{
(Color online)
Possible states in  the  $V_{\perp}^\mathrm{b}$-$T$ plane 
 with $t_{2}^*=0.0$ and $h=0.0$.
The solid line represents the boundary of the respective phase, and 
the dotted line is the energy, $t_{1}$, corresponding to interchain hopping. 
The dashed line denotes the spin gap $\Delta_{\sigma}$ obtained from 
eq. (\ref{eq:g_sigmaplus}).
}
\label{fig:enest0.0-E-vpb}
\end{figure}

Since  the RG equations in  the case of perfect nesting 
  reduce  to those obtained 
for two-coupled chains, \cite{Tsuchiizu2005} 
  the dominant state  in the absence of   magnetic field  
   is then  either SC$ d$ or CDW.
For  $V_{\perp}^\mathrm{b}=0$, 
the system is characterized by  a singlet   SC$ d$ state 
 as shown in Fig. \ref{fig:enest0.0-res} (a). 
Such a result is well known for  the Hubbard ladder model 
 with repulsive interactions \cite{Fabrizio1}.
The SDW correlations, which are the most dominant fluctuations
   at high temperature, becomes less dominant
 at low temperature due to the formation of   a spin gap. 
For $V_{\perp}^\mathrm{b}$ larger than a critical value, 
 $V_{\perp \mathrm{c}}^\mathrm{b}$, 
the CDW becomes in turn  the dominant state as illustrated
 in   Fig. \ref{fig:enest0.0-res} (b).
The  SC$f$  correlations are enhanced, 
 but show a weaker increase than CDW at low temperature due to the spin gap.

In Fig. \ref{fig:enest0.0-E-vpb}, 
 the most dominant states  are shown in  the
  $V_{\perp}^\mathrm{b}$-$T$ plane.   
At high temperature ($T>t_{1}$), corresponding to one-dimensional regime, 
  SDW is  dominant. 
 With decreasing temperature, the effect of interchain hoppings grows
  and   the dominant state  is in turn  
 either  SC$d$ or CDW. 
On a temperature scale, this occurs
 at a characteristic temperature given by the solid line. 
 This scale decreases and becomes zero at 
    $V_{\perp}^\mathrm{b} = V_{\perp \mathrm{c}}^\mathrm{b}(\simeq 0.83)$; it 
   increases monotonically 
   for   $V_{\perp}^\mathrm{b} > V_{\perp \mathrm{c}}^\mathrm{b}$. 
This  behavior  resembles to that of the spin gap, $\Delta_{\sigma}$, 
 as shown by the dashed line in Fig. \ref{fig:enest0.0-E-vpb}. 
 The existence of such a critical point of $V_{\perp \mathrm{c}}^\mathrm{b}$
  has been also shown   for two-coupled chains.
\cite{Tsuchiizu2005} 
A similar behavior shown by $ \Delta_\sigma$
 indicates that spin degrees of freedom
 are also critical at  $V_{\perp c}^\mathrm{b}$.
 The behavior around  the  quantum critical point, $V_{\perp c}^\mathrm{b}$,
 is ascribed to the competition between SC$d$ and CDW,
which are associated to different spin gaps.  
 The spin gap for SC$d$ is formed by  interchain pairing,
 whereas   that of CDW results from  intrachain interactions. 
Therefore
  $\Delta_\sigma$  vanishes at the critical point where 
the symmetry  of the gap changes.

\begin{figure}[tb]
\begin{center}
\includegraphics[width=7cm]{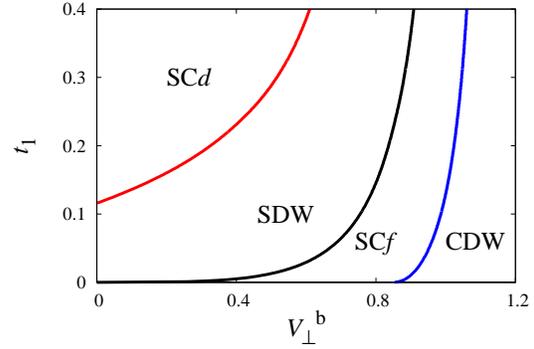}
\end{center}

\vspace*{-.3cm}
\caption{
(Color online)
Phase diagram in the  $V_{\perp}^\mathrm{b}$-$t_1$ plane   
 with $h = 0.0002$.
The solid lines designate the boundary for respective phases. 
The line  between SDW and SC$_f$ 
  refers to  $V_{\perp \mathrm{c}}^\mathrm{b}$.   
 }
\label{fig:p-h-vpb}
\end{figure}

 We now turn to  the effect of magnetic field, $h$, which brings  
 new states  due to  its influence on  the spin gap.  
When $\Delta_\sigma$ is destroyed by the magnetic field, 
 the magnetic state is expected to be either the transverse SDW or  SC$f$. 
 Such a region actually exists 
 when $V_{\perp}^\mathrm{b}\sim V_{\perp \mathrm{c}}^\mathrm{b}$. 
 The phase diagram in the  $V_{\perp}^\mathrm{b}$ - $t_1$
 plane is shown in  Fig. \ref{fig:p-h-vpb}.
  The boundary between SC$d$ and SDW 
 (and also between SC$f$ and CDW)
 is estimated from the condition, 
$\Delta_\sigma   \rightarrow 0$.
It is found that the effect of $h$ on   SC$d$ is larger than  it is for CDW. 
 The SDW state 
 thus takes place for 
 $V_{\perp}^\mathrm{b}< V_{\perp \mathrm{c}}^\mathrm{b}$,
  whereas the SC$f$ state appears for 
  $V_{\perp}^\mathrm{b} > V_{\perp \mathrm{c}}^\mathrm{b}$. 
  Moreover, $V_{\perp \mathrm{c}}^\mathrm{b}$
    is 
a critical value
 for  
 the crossover between SDW and SC$f$  as the dominant fluctuations. 
 In our two-loop approach, the fact that both  SDW and SC$f$ 
 of   Fig. \ref{fig:p-h-vpb} vanish for $h=0$  
 is at variance with the emergence of SDW
 state found in  ref. \citen{Abramovici1}   at low temperature 
 using one-loop RG. 

\subsection{Nesting deviations and superconductivity}

\begin{figure}[t]
\begin{center}
\includegraphics[width=6.5cm]{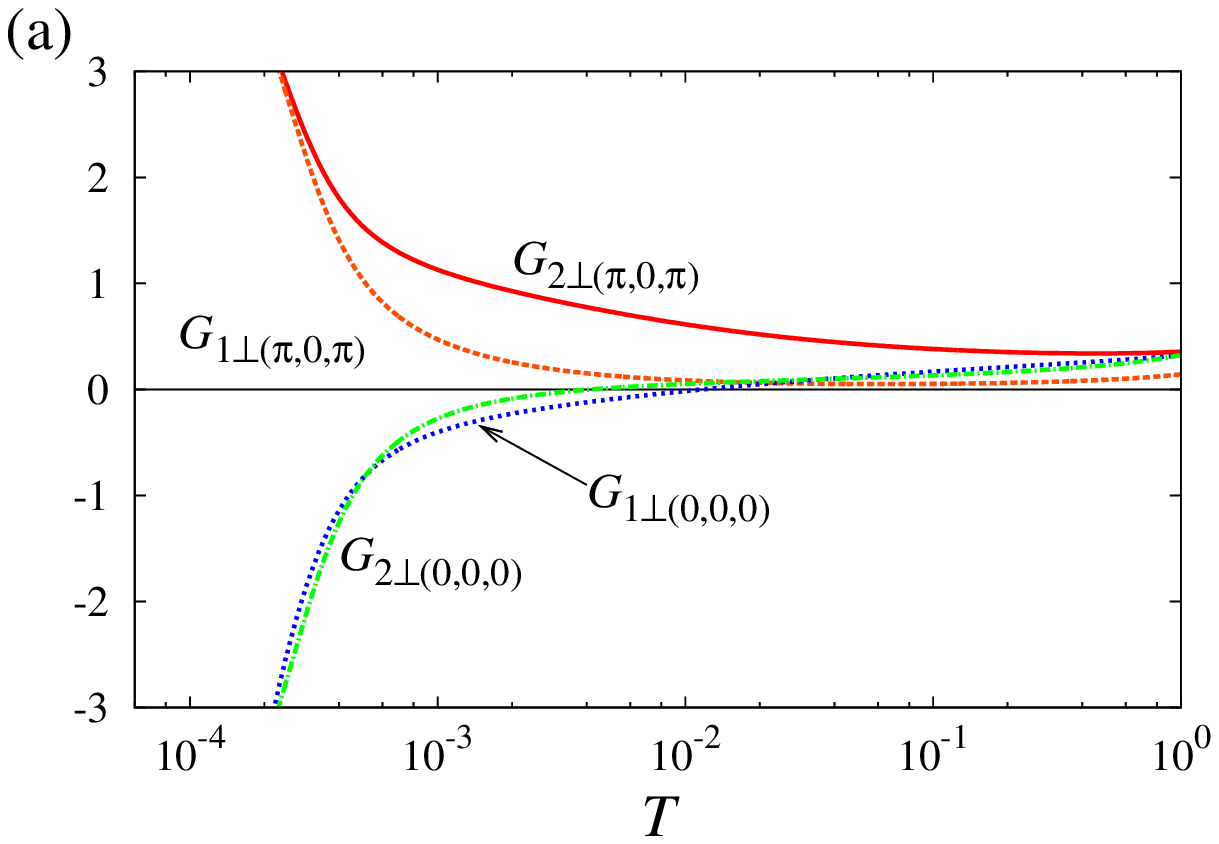}
\includegraphics[width=6.5cm]{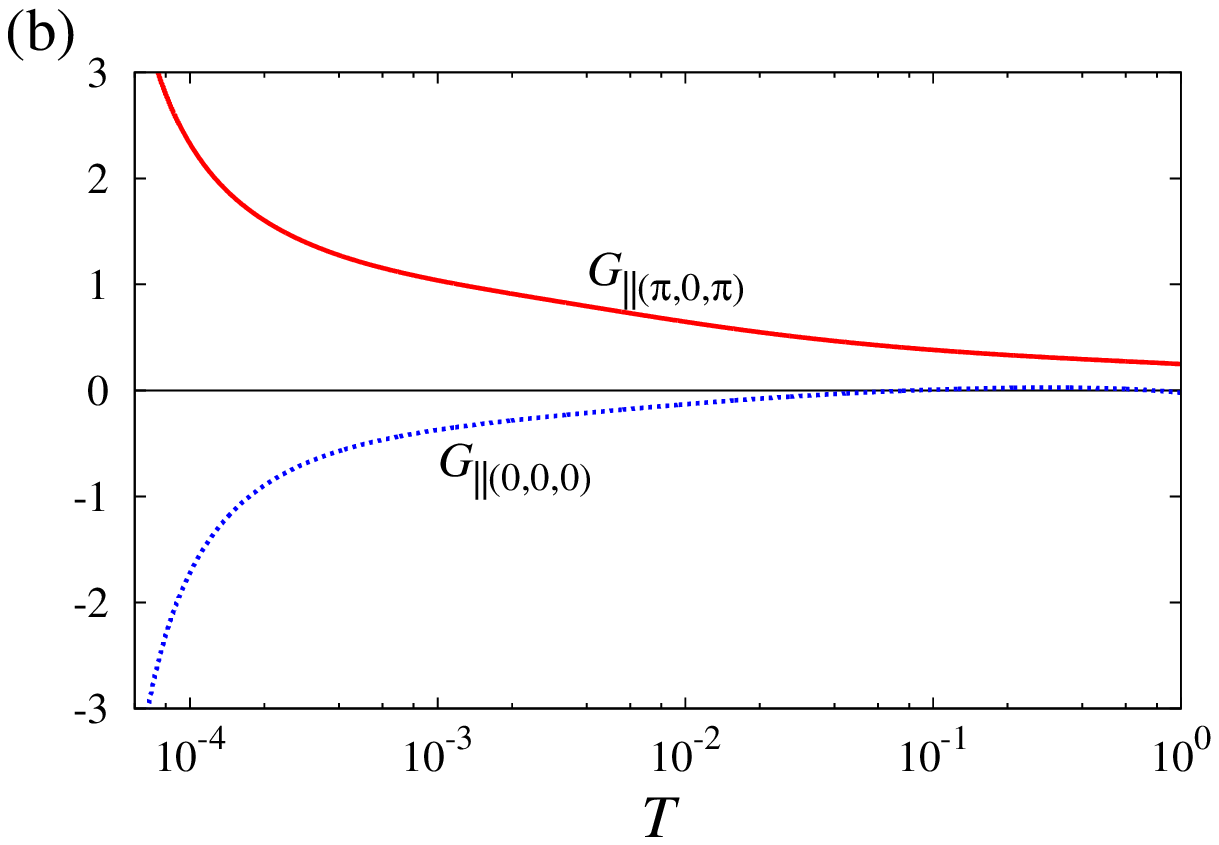}
\includegraphics[width=6.5cm]{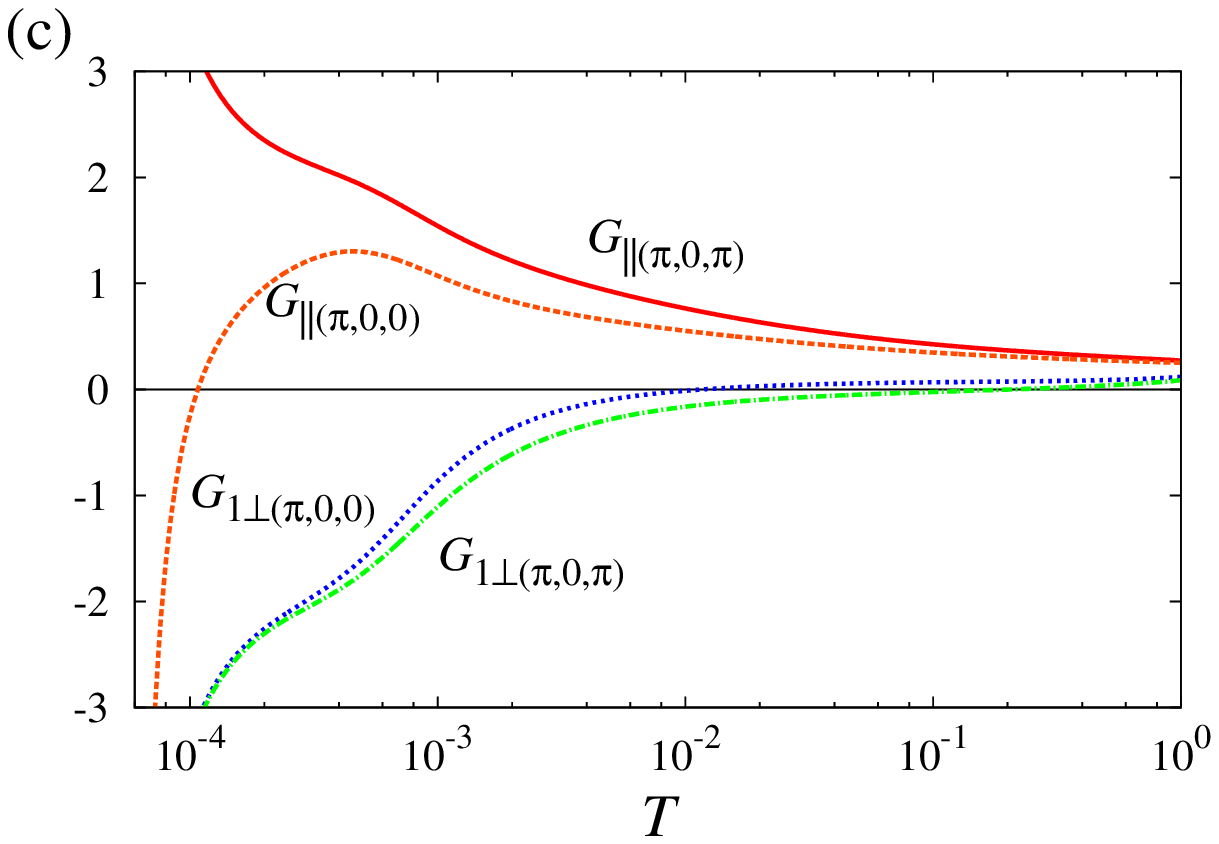}
\end{center}

\vspace*{-.3cm}
\caption{
(Color online)
Temperature dependence of the coupling constants for  
 $t_{2}^*/t_{1}=0.001$ and $h=0.0$,
with fixed $V_{\perp}^\mathrm{b}=0.7$ (a), $0.88$ (b), and $1.0$ (c).
Only the relevant coupling constants are shown.
}
\label{fig:imp-g}
\end{figure}
\begin{figure}[t]
\begin{center}
\includegraphics[width=6.5cm]{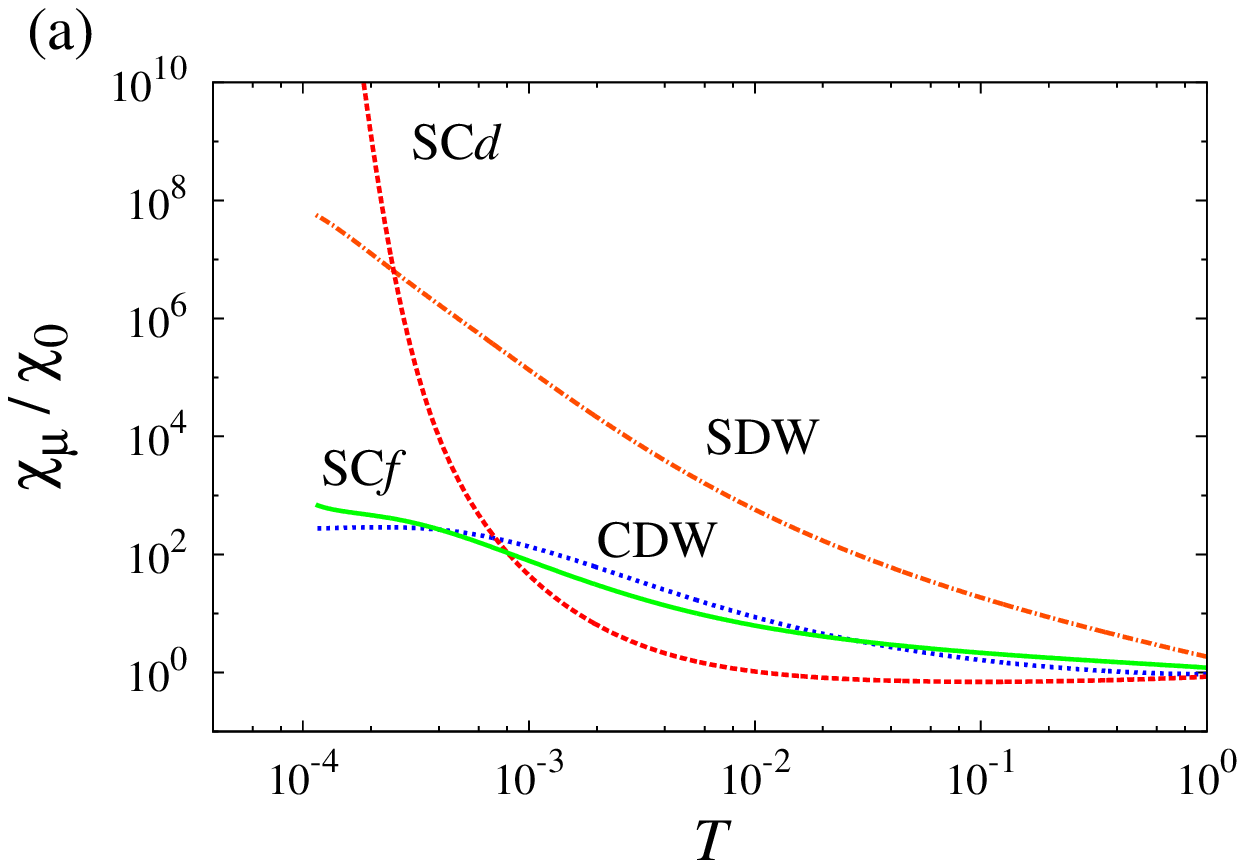}
\includegraphics[width=6.5cm]{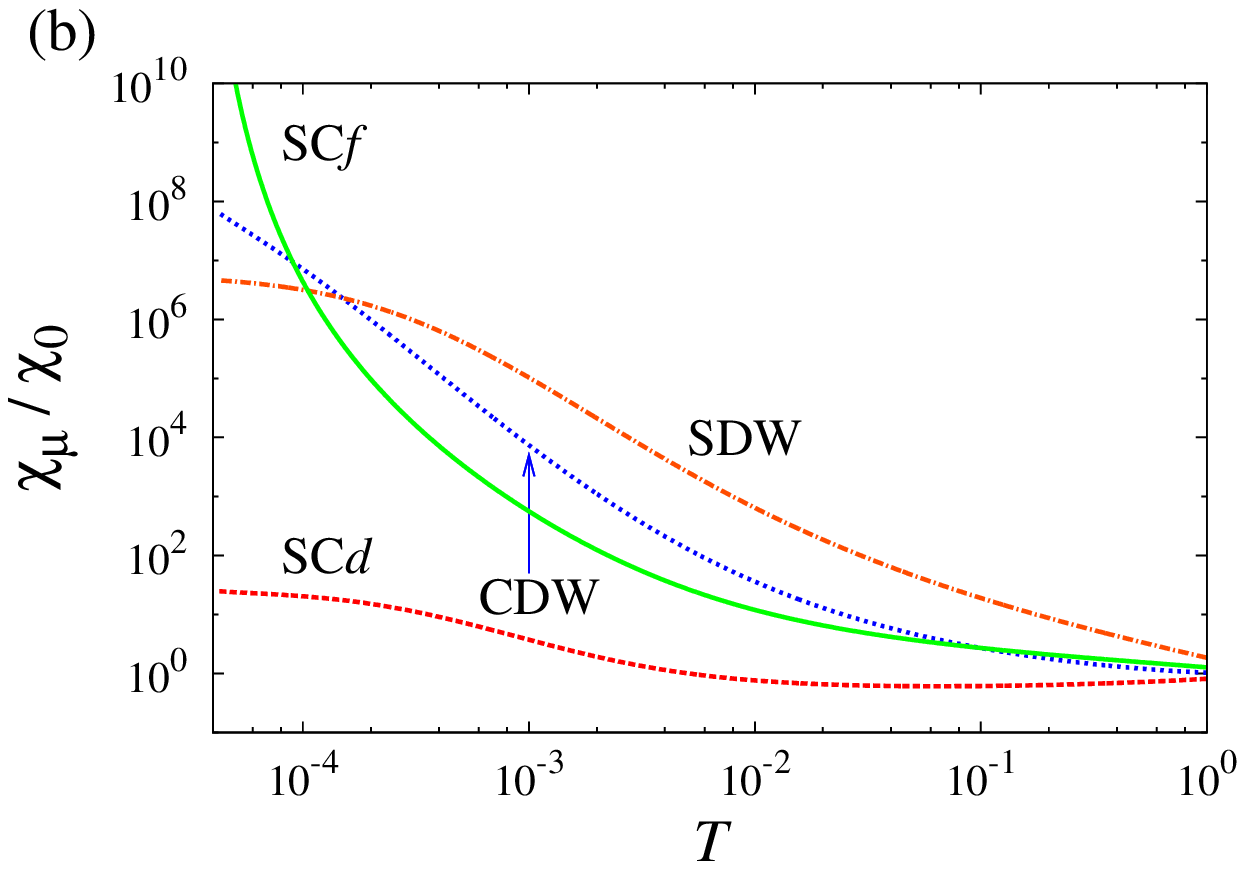}
\includegraphics[width=6.5cm]{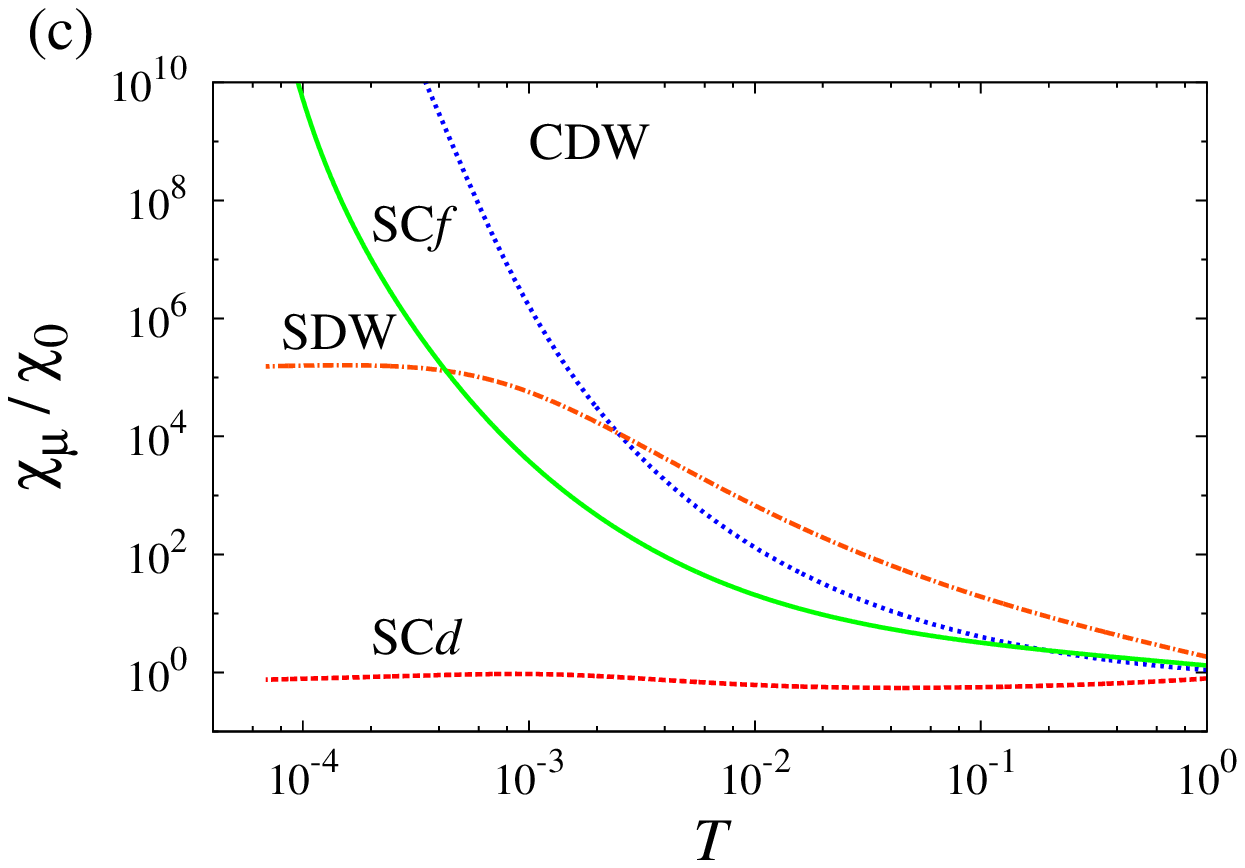}
\end{center}

\vspace*{-.3cm}
\caption{
(Color online)
Temperature dependence of response functions at
 $t_{2}^*/t_{1}=0.001$ and $h=0.0$. 
with fixed $V_{\perp}^\mathrm{b}=0.7$ (a), $0.88$ (b), and $1.0$ (c).
}
\label{fig:imp-res}
\end{figure}

We now examine the SC state in the presence of  nesting deviations, which is the 
 main subject of the present paper. 
 We  look at  the possible states at $t_2^*\ne 0$ and  low temperature 
   by choosing 
 $V_{\perp}^\mathrm{b}=0.7$ (a), $0.88$ (b), and $1.0$ (c),
 namely for small, intermediate, and large interchain couplings.
Figures  \ref{fig:imp-g} (a), (b), and (c) 
show  the temperature dependence of    coupling constants which 
 give rise to  the response functions for SC$d$, SC$f$, and CDW respectively.
 The other couplings  
(not shown in the figures)
 only contribute 
a lesser degree  to the response functions.
These functions   are traced  in 
 Figs. \ref{fig:imp-res} (a)-(c).

For  $V_{\perp}^\mathrm{b} =0.7$ [Fig. \ref{fig:imp-g} (a)], 
 the values of  $G_{{1\perp}(\pi,0,\pi)} $ 
      and  $G_{{2\perp}(\pi,0,\pi)}$ at the fixed points
 are  positive, while  those  of 
 $G_{{1\perp}(0,0,0)}$ and  $G_{{2\perp}(0,0,0)}$ are negative.
 The SC$d$ response function is then the  strongest
  and becomes  the dominant state. 
Since  the coupling constant $G_{{2\perp}(0,0,0)}$ 
 changes its sign and becomes relevant, 
 a spin gap appears   
 [see eq. (\ref{eq:g_sigmaplus})],   which is    crucial to SC$d$.  
As for the relevant  coupling $G_{{2\perp}(\pi,0,\pi)} (>0)$, it enhances 
 both SDW and SC$d$,
as seen from eqs. (\ref{eq:dlnz-SDW}) and (\ref{eq:dlnz-SCd}).
Figure \ref{fig:imp-res} (a) shows the temperature dependence  of the  
  response functions, where 
 the SDW state is found to be the dominant state at high temperature,  but  
   becomes  sub-dominant  at low temperature.   
The amplitude of  SDW correlations is reduced by  nesting deviations. 
Thus it is found that 
 SC$d$ pairing is induced by spin fluctuations with the coupling 
 $G_{{2\perp}(\pi,0,\pi)} (>0)$.

For  $V_{\perp}^\mathrm{b} =0.88$ [Fig. \ref{fig:imp-g} (b)], 
the relevant  couplings are 
given by   $G_{{\|}(\pi,0,\pi)} (>0)$ and  $G_{{\|}(0,0,0)} (<0)$ 
 which are quite different from those 
obtained in  (a) for weaker $V_\perp^\mathrm{b}$; 
   SC$f$   fluctuations are then dominant, 
as seen from  eq. (\ref{eq:dlnz-SCf}). 
The spin gap vanishes
   since the combination of couplings given by 
  eq. (\ref{eq:g_sigmaplus}) remains positive. 
Thus if  $\Delta_\sigma=0$,
  the dominant contribution to the SC coupling comes from   
 density  fluctuations
 for which the relevant coupling with parallel spins, namely
 $G_{{\|}(\pi,0,\pi)} (>0)$,
  is connected  to   charge fluctuations. 
It is worth noting that  long wave length  spin fluctuations     also 
 promote the  SC$f$ state 
since that  $G_{{\|}(0,0,0)} <0$  strengthens  
  SC$f$ correlations  and  also the  uniform spin susceptibility 
[eq. (\ref{eq:uniform_susceptibilty})], as it will be discussed later.
The  CDW fluctuations are well  developed  at temperatures 
 just above the region where SC$f$ is dominant 
  [Fig. \ref{fig:imp-res} (b)]. 
 At these temperatures, 
 the CDW response function    is  nearly the same as that of SDW, 
    implying  a    coexistence of 
  spin and   charge fluctuations 
   for  intermediate 
    $V_{\perp}^\mathrm{b}$.
We finally note that the dominance of  SC$f$ state is 
obtained in the presence of 
  nesting deviations ($t_{2}^* \not= 0$), which  
  suppresses the divergence of  CDW in the  low temperature limit.

For  $V_{\perp}^\mathrm{b} =1.0$ [Fig. \ref{fig:imp-g} (c)],  
the relevant  couplings are 
   $G_{{\|}(\pi,0,0)} (>0)$,  $G_{{\|}(\pi,0,\pi)} (>0)$,
  $G_{{1\perp}(\pi,0,0)} (<0)$, and   $G_{{1\perp}(\pi,0,\pi)} (<0)$,
 which lead  to a CDW state as seen from eq. (\ref{eq:dlnz-CDW}). 
 The behavior of   $G_{{\|}(\pi,0,\pi)} (>0)$ is similar to that of (b),
 and then  a similarity of  CDW and SC$f$ response functions 
 is expected. 
However, the nature of the spin gap as 
 induced by  $G_{{1\perp}(\pi,0,0)} (<0)$ differs.
 For large $V_{\perp}^\mathrm{b}$, it is expected that
  the dominant state is  CDW, whereas SC$f$
 is sub-dominant. 
Here the difference in    sign for  $G_{{1\perp}(\pi,0,\pi)}$ 
in comparison to the case (a) is essential 
     to the predominance   of either SC$d$ or CDW.

\begin{figure}[t]
\begin{center}
\includegraphics[width=7cm]{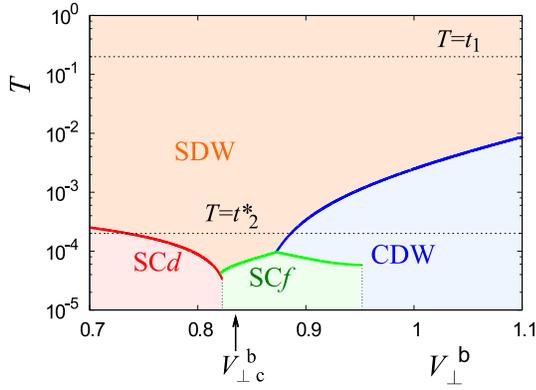}
\end{center}

\vspace*{-.3cm}
\caption{
(Color online)
The $T-V_{\perp}^\mathrm{b}$ "phase diagram" with $t_{2}^*/t_{1}=0.001$ 
and 
$h=0.0$.
}
\label{fig:enest0.001-E-vpb}
\end{figure}

 The dominant states are summarized  in the  $V_{\perp}^\mathrm{b}$-$T$
  phase diagram  of Fig. \ref{fig:enest0.001-E-vpb}.
This phase diagram shows some  similarity  with the one of 
    Fig.~\ref{fig:enest0.0-E-vpb}
in the sense that the  SDW state is found 
at high temperature, namely for     $T \gtrsim t_{2}^*$. 
However, for $T \lesssim t_{2}^*$,
 the SC$f$ state  comes in between 
the regions of    the SC$d$ and CDW states.  
In the zone where 
 $T\lesssim t_{2}^*$
  and $V_{\perp}^\mathrm{b} \gtrsim V_{\perp \mathrm{c}}^\mathrm{b}$, 
the  CDW state is dominant and the SC$f$ state
  is sub-dominant, 
 but the CDW state is suppressed and the SC$f$ state becomes dominant 
   for  $t_{2}^* \not= 0$ --  a result well explained by the fact that 
  nesting deviations have  a   detrimental  influence primarily 
    on   density-wave correlations.
  The effect of $t_{2}^*$, when 
    $V_{\perp}^\mathrm{b} < V_{\perp c}^\mathrm{b}$, is small, 
     because
   the   SC$d$ state 
  is less affected by  nesting deviations in that region. 

\begin{figure}[b]
\begin{center}
\includegraphics[width=7cm]{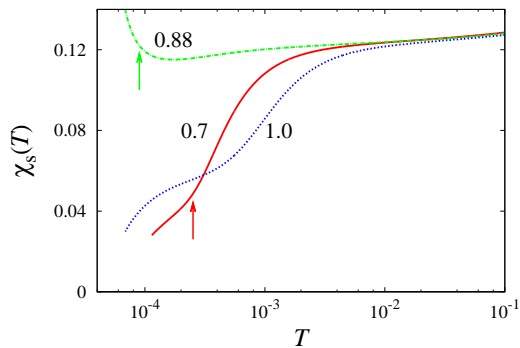}
\end{center}

\vspace*{-.3cm}
\caption{
(Color online)
Temperature dependence of the uniform susceptibility 
 for $V_{\perp}^\mathrm{b}= 0.7$, $0.88$, and $1.0$, 
 where the parameters are the same as 
 Figs. \ref{fig:imp-res} and \ref{fig:imp-g}.
The arrow denotes the temperature at which the SC state  becomes dominant with 
decreasing temperature. 
}
\label{fig:chi_s}
\end{figure}

 At this point we would like to comment   on the behavior of the
uniform magnetic susceptibility, $\chi_s$,  given by
eq. (\ref{eq:uniform_susceptibilty}).
In Fig. \ref{fig:chi_s}, the temperature dependence of this quantity 
 is shown  for the $V_{\perp}^\mathrm{b}= 0.7$, $0.88$, and $1.0$ 
 cases considered above,
 which correspond to the  SC$d$, SC$f$, and  CDW states, respectively.
It should be noticed that  $\chi_s$ increases  when
the system is entering in   the  SC${f}$ state, indicating  an
enhancement of long wave length spin correlations.  
This contrasts  with the cases where   SC$d$ and CDW states prevail 
 and $\chi_s$  decreases rapidly 
 due to the formation of a spin gap.

\begin{figure}[t]
\begin{center}
\includegraphics[width=7cm]{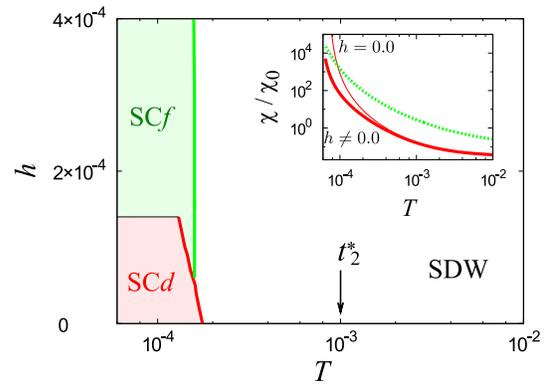}
\end{center}

\vspace*{-.3cm}
\caption{
(Color online)
Phase diagram of the most dominant states 
 in the $T$-$h$ plane with  $V_{\perp}^\mathrm{b}=0.78$ and 
 $t_{2}^*/t_{1}=0.005$. 
In the inset, the temperature dependence of the SC$d$ (solid line) 
 and SC$f$ (dotted line) response functions 
 is shown where the  thin (bold) solid line  corresponds 
 to the case for $h=0$ $(0.0002)$.
}
\label{fig:enest0.001-h-E}
\end{figure}

Now we study the effect of a magnetic field on the SC$d$ state for 
 the intermediate   interchain interaction, namely  in the region where  
 the sub-dominant SC$f$ state is close to SC$d$. 
Figure \ref{fig:enest0.001-h-E} 
 shows the $T$ \textit{vs} $h$ phase diagram 
 for the most dominant states at  $V_{\perp}^\mathrm{b}=0.78$. 
 At low temperature, there is  a crossover   from  
 SC$d$   to the SC$f$ state due to the suppression of the  spin gap with $h$.  
In the inset, the temperature dependence  of the response functions for 
 SC$d$ and SC$f$ are shown at finite ($h$= 0.0002)
and zero magnetic field.
The SC$d$ correlations are strongly suppressed by $h$, while 
the magnetic field has essentially no influence  on     SC$ f$ correlations.

\section{Summary and Discussion}

 We have examined by  the  two-loop  RG method the singlet SC$d$ 
 and the triplet SC$f$  superconducting  states
 in the framework of a 4-patch model with nesting deviations. 
The triplet SC$f$ state, which  is absent 
in the model for \textit{intrachain} interactions only,
is found to develop from the combined effect of 
  \textit{interchain} repulsive interactions and
   nesting deviations.

\begin{figure*}[t]
\begin{center}
\includegraphics[width=7cm]{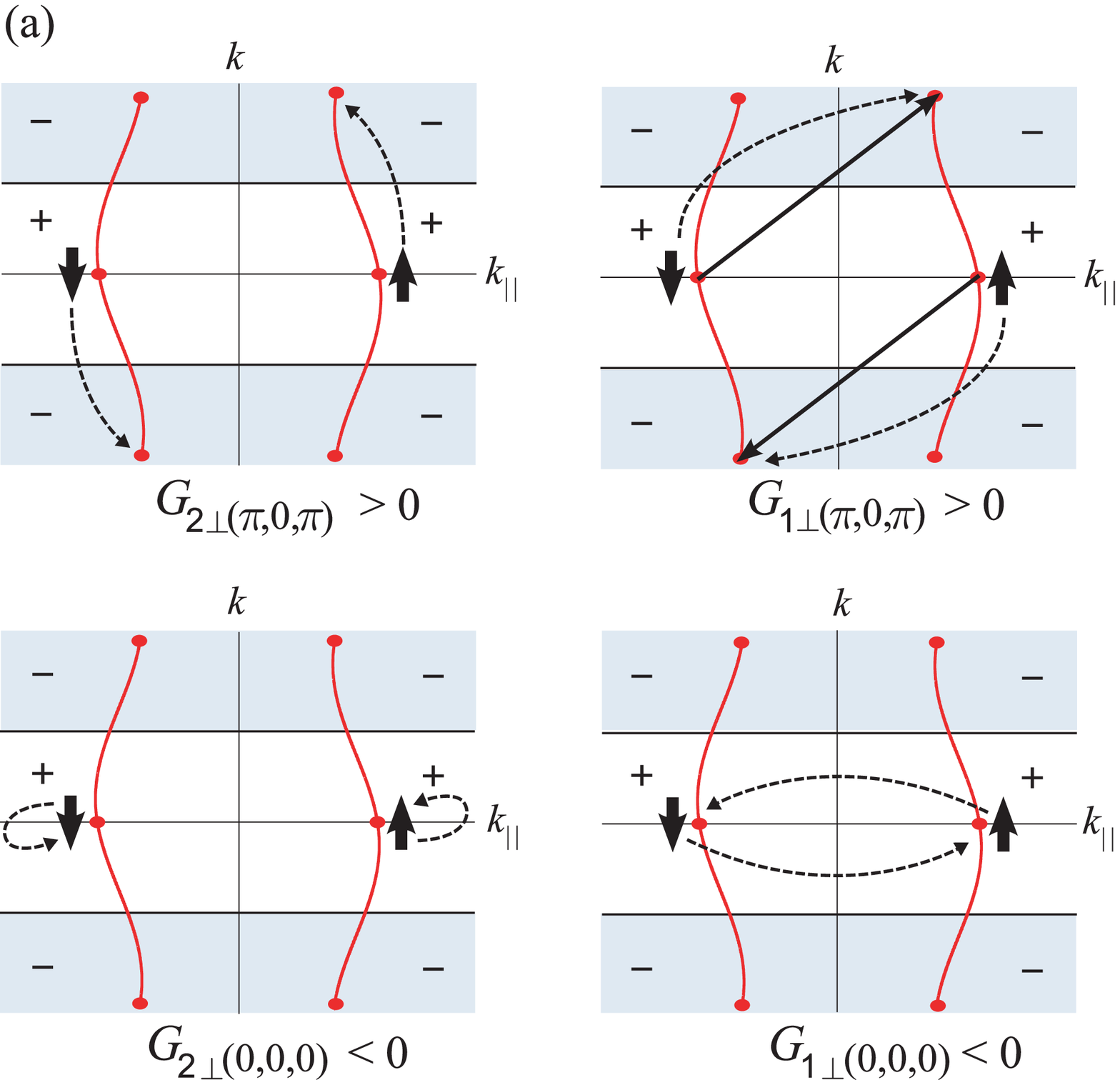}
\hspace*{1cm}
\includegraphics[width=7cm]{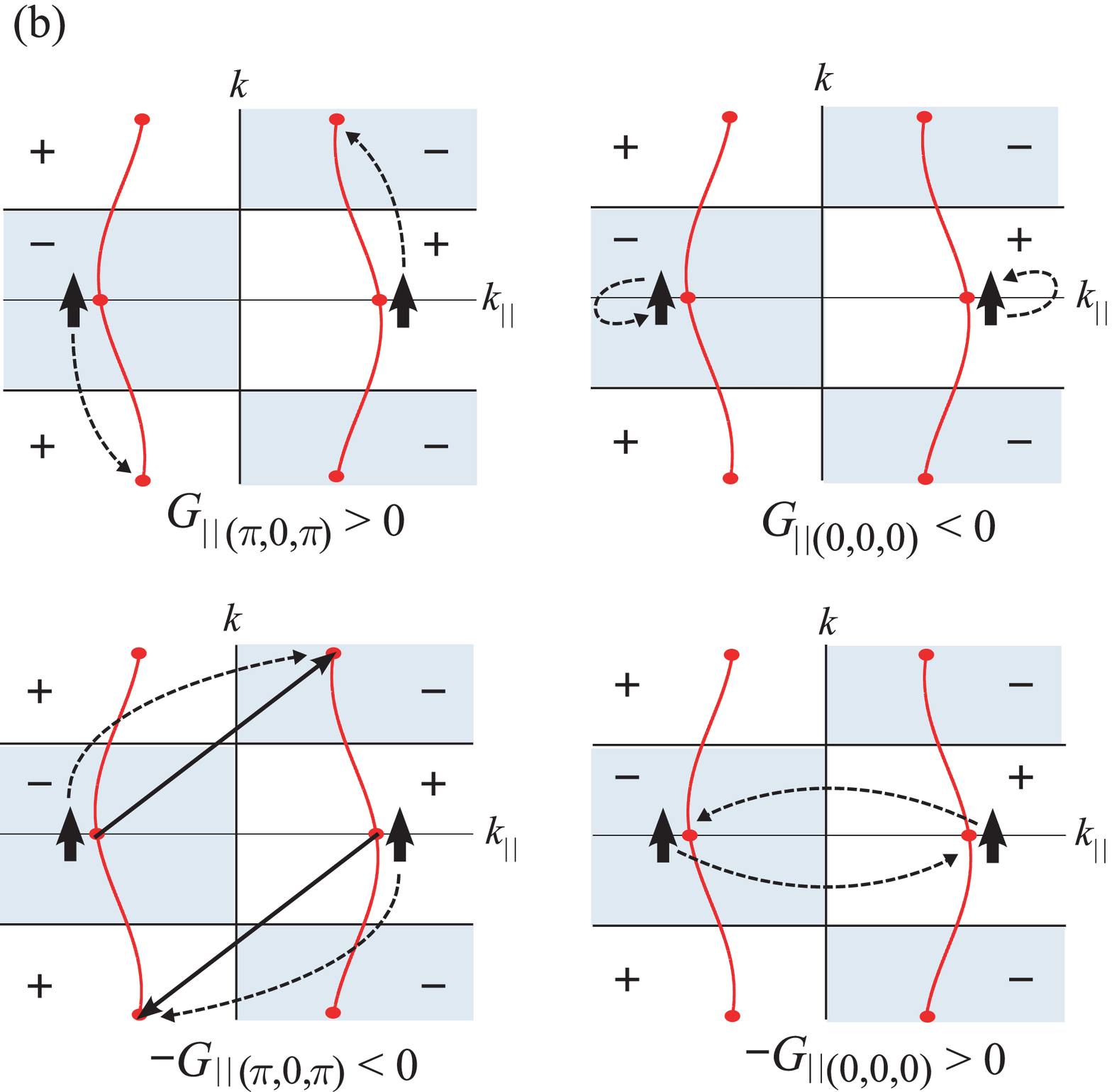}
\end{center}

\vspace*{-.3cm}
\caption{
(Color online)
Scattering processes   giving rise to  
 the SC$d$ state (a) and SC$f$ state (b). 
The Fermi surface is given by the line connecting 
 the Fermi points (closed circle).
}
\label{fig:interpre}
\end{figure*}

    It is of importance to return to the mechanism
 of formation of   the superconducting state 
  and examine  the scattering processes 
that are pertinent to this state.  
This is shown in Fig. \ref{fig:interpre}, 
where the symbol  $\pm$ in the  $k_{\|}$-$k$ plane 
 denotes the sign of the  superconducting gap function for  the 
SC$d$  [i.e.,  $\cos k$ in eq. (\ref{O-SCd}) (a)] and SC$f$ 
 [i.e., $\sin k_{\|} \cos k$ 
 in eq. (\ref{O-SCd}) (b)] phases.  
The scattering processes  for parallel and anti-parallel spins
  are described by a dashed arrow, while  
the long continuous  arrows stand for the nesting vector. 
The sign  of the renormalized   coupling constants  are also 
 stated in each case considered.    
A sign change of the gap function following the scattering 
 occurs for   positive renormalized coupling constants,
 while the sign remains the same for attractive ones.
Thus all the scattering processes shown in Fig. \ref{fig:interpre} 
 gain the energy.

Figure \ref{fig:interpre} (a) depicts
   scattering processes with anti-parallel spins  leading to the SC$d$ state.
The interaction with interchain momentum transfer $\pi$, i.e., 
the interband scatterings,  give  rise to the repulsive couplings  
among which $G_{2\perp (\pi,0,\pi)}$ 
is the dominant contribution and  
  $G_{1\perp (\pi,0,\pi)}$   also becomes relevant,  
     leading to the  growth of  spin fluctuations 
 as explained  after eq. (\ref{eq:Ih}).
The sign of the gap function changes in these scattering process.
For the  couplings  $G_{2 \perp(0,0,0)}$ and 
 $G_{1 \perp(0,0,0)}$, which 
  become attractive through the renormalization, 
   the  sign of the gap function remains the same in the scattering process.

We now turn to  the SC$f$ state, where
 the relation between the sign change of the gap function and 
 the forward scattering process with  parallel spins  
  is shown in Fig. \ref{fig:interpre} (b) (upper panel). 
 For the repulsive $G_{\|(\pi,0,\pi)}$,  
    the  gap function  is opposite in sign,  
  whereas the sign is the same for the attractive   $G_{\|(0,0,0)}$. 
  The backward scattering  shown in the lower (b) panel    
 also  favors the SC$f$ state, since  the signs of  
 the respective coupling constants, 
 $- G_{\|(\pi,0,\pi)}<0$ and $- G_{\|(0,0,0)}>0$, 
 are consistent  with 
 the change of the sign in the gap function.

\begin{figure}[t]
\begin{center}
\includegraphics[width=7cm]{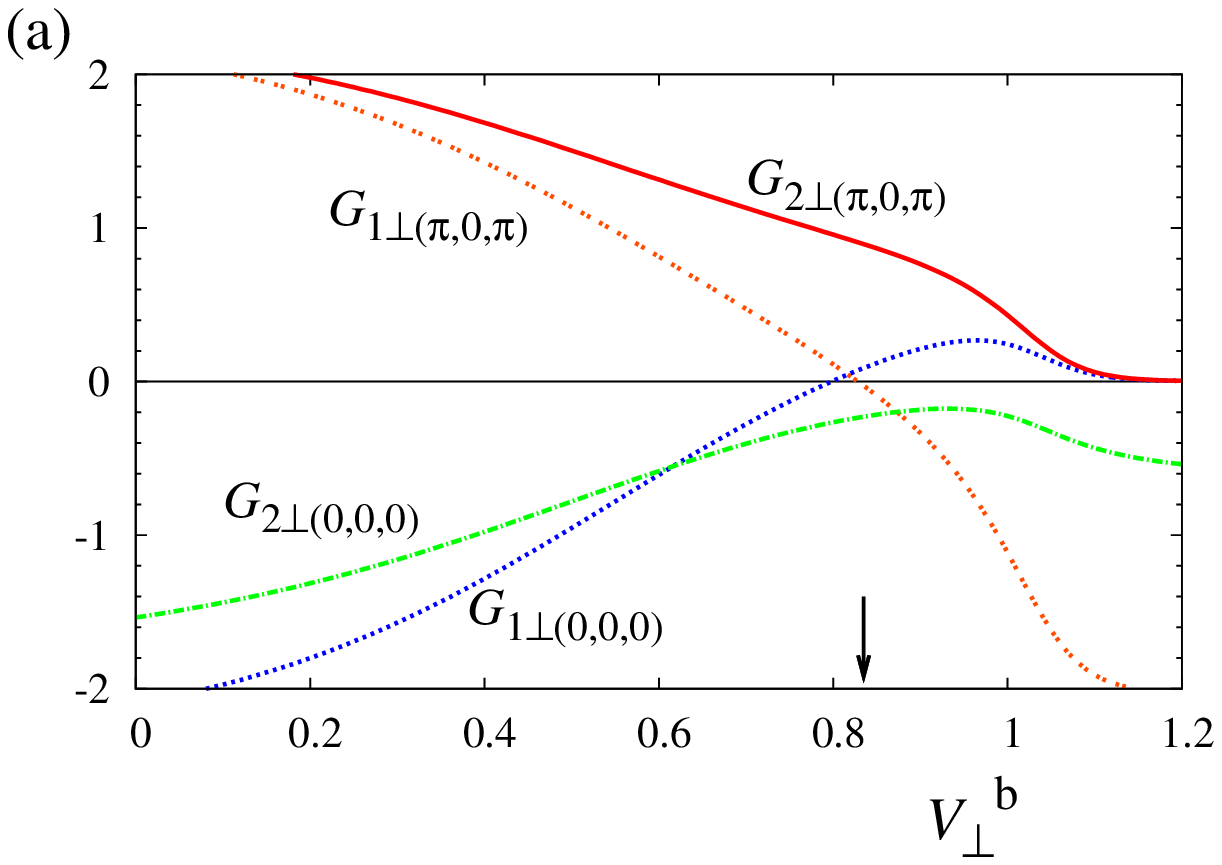}
\includegraphics[width=7cm]{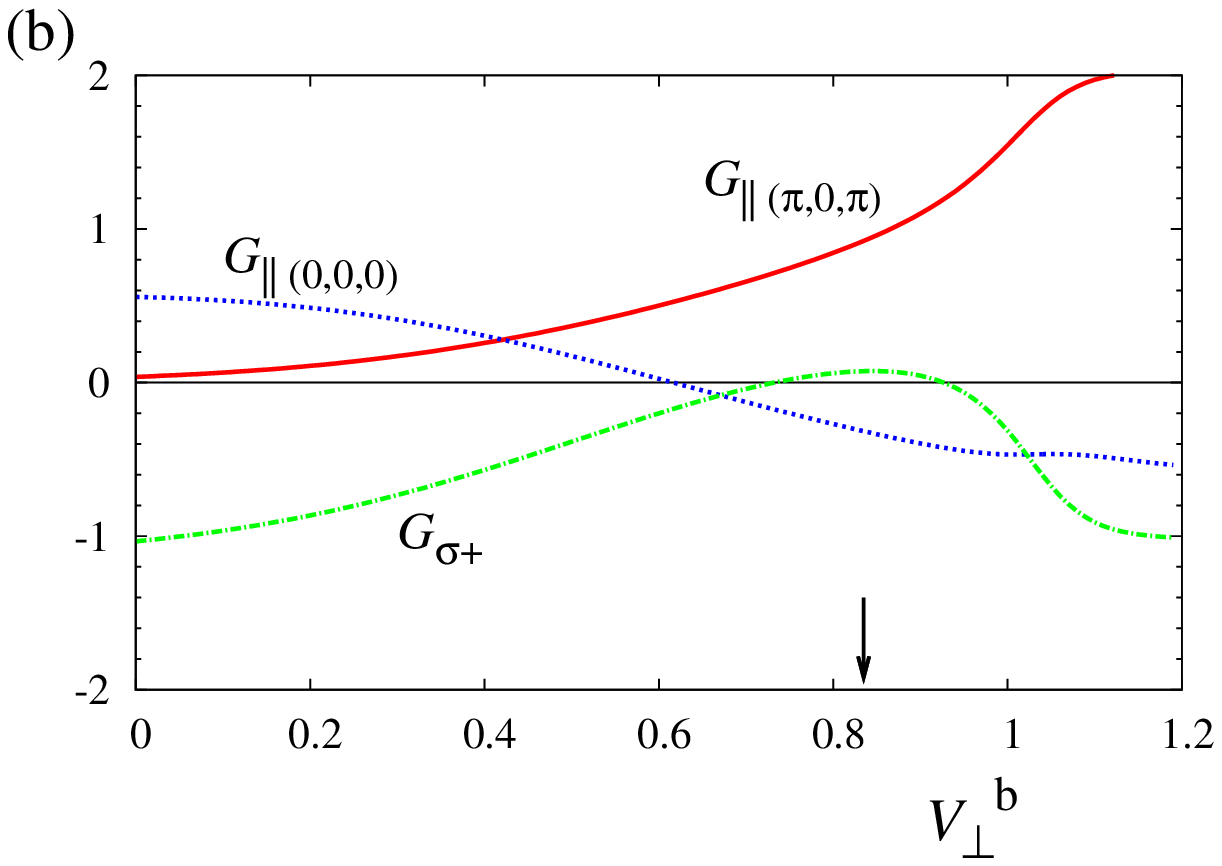}
\end{center}

\vspace*{-.3cm}
\caption{
(Color online)
The $V_{\perp}^\mathrm{b}$ dependence of the renormalized 
 coupling constants at $T=10^{-3}$ for the SC$d$ (a) 
  and  SC$f$ (b) states, and for $G_{\sigma +}$. Here  $t_{2}^*=0.001t_{1}$ and $h=0.0$.
The initial values for 
 $G_{1 \perp(0,0,0)}$,  $G_{1 \perp(\pi,0,\pi)}$, 
 $G_{2 \perp(0,0,0)}$,  $G_{2 \perp(\pi,0,\pi)}$, 
 $G_{\|(0,0,0)}$ and  $G_{\| (\pi,0,\pi)}$, 
  are respectively given by 
 0.45, 0.45, 0.45, 0.45, 0.00 and 0.00 for $V_{\perp}^\mathrm{b} = 0$ and 
  0.72, 0.18, 0.45, 0.45, -0.27 and 0.27    for $V_{\perp}^\mathrm{b} =1.2$.  
 The arrow represents $V_{\perp c}^\mathrm{b}$
 }
\label{fig:imp-g-vpb}
\end{figure}

We now comment on the crossover from the SC$d$ to SC$f$ states
 as a function of $V_{\perp}^\mathrm{b}$. 
 The $V_{\perp}^\mathrm{b}$ dependence 
 of relevant coupling constants at low temperature ($T=10^{-3}$)
   is shown in Figs. \ref{fig:imp-g-vpb} (a) and (b), which 
    correspond to the SC$d$ and SC$f$ cases, respectively. 
 For the SC$d$ state, 
the coupling $G_{1\perp(0,0,0)}$  evolves from  negative 
   to   positive values, suggesting a vanishing   
    spin gap. 
The resultant reduction of the SC$d$ state
 by $V_{\perp}^\mathrm{b}$ is reasonable 
 because  
 the interchain  spin singlet state  is  
  destroyed by 
 the formation of  CDW state for  moderate $V_{\perp}^\mathrm{b}$. 
With increasing $V_{\perp}^\mathrm{b}$, the repulsive interactions 
$G_{2\perp(\pi,0,\pi)}$ and  $G_{1\perp(\pi,0,\pi)}$ decrease 
 and the  amplitude of the SC$d$ correlation  is reduced. 

In Fig. \ref{fig:imp-g-vpb} (b),
  the coupling $G_{\|(\pi,0,\pi)}$ increases from zero, while 
the coupling $G_{\|(0,0,0)}$ decreases to negative value.
Negative $G_{\|(0,0,0)}$ indicates the absence of  spin  gap. 
These features yield in turn the development of
  the  SC$f$ state,  as a consequence of   charge fluctuations
 and nesting deviations. 
 From Fig. \ref{fig:imp-g-vpb} (b), 
 the $V_{\perp}^\mathrm{b}$ interval, where $G_{\sigma +} > 0$  ,
  suggests that  the   spin  gap vanishes 
    for 
      $V_{\perp}^\mathrm{b} \sim V_{\perp c}^\mathrm{b}$. 
However,
it is found that the SC$f$ state moves to the CDW  state 
by noting that  $G_{\sigma +} > 0$ for  
$V_{\perp}^\mathrm{b} =  0.88$, 
and   $G_{\sigma +} < 0$ for $V_{\perp}^\mathrm{b} = 0.92$.

\begin{figure}[t]
\begin{center}
\includegraphics[width=6cm]{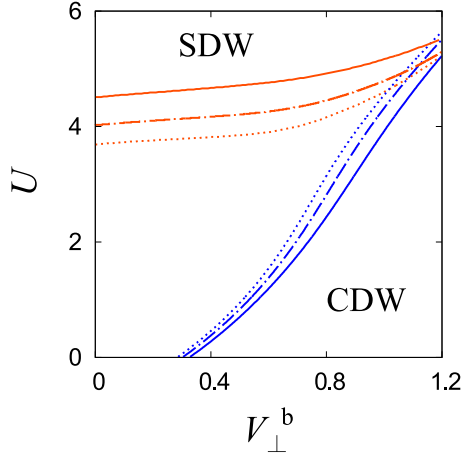}
\end{center}

\vspace*{-.3cm}
\caption{
(Color online)
The regions of SDW and CDW dominant correlations 
 in the $V_{\perp}^\mathrm{b}$-$U$ plane
defined by the condition 
$\chi_\mathrm{SDW} (\chi_\mathrm{CDW})> 10^6\chi_{0}$  
  at the fixed temperatures  $T =2 \times 10^{-4}$ (dashed line), 
$4 \times 10^{-4}$ (dash-dotted)  and 
$8 \times  10^{-4}$ (solid line). Here 
 $t_2^* = 2 \times 10^{-4}$. 
In the remaining in between region   
$\chi_\mathrm{SDW} (\chi_\mathrm{ CDW}) < 10^6\chi_{0}$.
}
\label{fig:u-vpb}
\end{figure}

Finally, we comment on the effect of $V_{\perp}^\mathrm{b}$ on the SDW state. 
Within the conventional treatment of RPA,
\cite{Scalapino1987,Schimahara1989,Tanaka2004} 
 the interference effect  between the scattering at 
  different transverse momenta is neglected. 
For example, 
in eq. (\ref{eq:coupling_initial_2}), which is relevant to 
  the SDW state, 
 the onsite-repulsion  $U$ 
is retained but the effect of $V_{\perp}^\mathrm{b}$ is strongly 
 reduced due to the summation  of 
  $k_1 = \pm \pi$ and $k_2 = \pm \pi$. 
However, the present RG  method  shows a clear 
 effect of  $V_{\perp}^\mathrm{b}$ on SDW 
   as illustrated by the strength of density-wave correlations 
in the   $V_{\perp}^\mathrm{b}-U$ plane at fixed $T$ (Fig. \ref{fig:u-vpb}). 
The SDW and CDW regions are defined by the corresponding response functions
 that become larger than the bare value $\chi_0$ 
at $l=0$ by a factor greater than $10^6$.
 The domain that separates  the two regions shows a much 
reduced amplitude of the density-wave response functions,
 while its area reduces by  
the decrease of temperature. 
The region for SDW is suppressed with increasing 
 $V_{\perp}^\mathrm{b}$ due to the increase of the charge fluctuations.

\section*{Acknowledgments}
The present research was partially supported by
Grant-in-Aid for Scientific Research on Innovative Areas 
20110002.

\appendix
\section{Two-loop level renormalization group}

The renormalization of the interchain hopping $t_1$ 
is given by
\cite{Tsuchiizu1} 

\begin{align}
\frac{d}{dl}t_{1}(l)=t_{1}(l)-\frac{1}{2N^2_{\perp}}\sum_{q,k,k'}G_{{\Sigma}_n(q,k,k')}J_{0(q,k,k')},
\end{align}
where the second term of r.h.s. comes from interactions.
Such an effect is negligibly small for the quarter-filled band 
 although the strong effect 
is expected due to the Umklapp scattering for the half-filled band,
 e.g., the confinement of the interchain  hopping.  
 \cite{SuzumuraPRB,TsuchiizuPRB}

We show the results at  the two-loop level 
 in eq. (\ref{full:RG}).  
Applying the calculations already obtained at    perfect nesting and  half-filling~\cite{Tsuchiizu1} 
to the case of the nesting deviation at quarter-filling,
 the self-energy   and   vertex corrections  are 
  $\Xi^{(2)}_{\Sigma(q,k_1,k_2,q',k')}$, $\Xi^{(2)}_{\nu (q,k_1,k_2,q',k')}$,
  which read  
\begin{subequations}
\begin{align}
&\Xi^{(2)}_{\Sigma(q,k_1,k_2,q',k')}=G_{{\Sigma}_n(q',k_1,k')}^2J_{1(q',k_1,k')}\nonumber\\
&\qquad
+G_{{\Sigma}_n(q',k_2,k')}^2J_{1(q',k_2,k')}+G_{{\Sigma}_n(q',-k_1+q,k')}^2J_{1(q',-k_1+q,k')}\nonumber\\
&\qquad
+G_{{\Sigma}_n(q',-k_2+q,k')}^2J_{1(q',-k_2+q,k')},
\\
&\Xi^{(2)}_{1\perp (q,k_1,k_2,q',k')}=J_{2(q+k';k_1,k_2;k',k'-q')}\nonumber\\
&\qquad
\times\bigl[G_{1\perp(q+q',k_1,k_2)}G_{2\perp(q-k_2+k',k'-q',k')}G_{\|(q-k_1+k',k',k'-q')}\nonumber\\
&\qquad
+G_{1\perp(q+q',k_1,k_2)}G_{\|(q-k_2+k',k'-q',k')}G_{2\perp(q-k_1+k',k',k'-q')}\bigl]\nonumber\\
&\qquad
+J_{2(-k';-k_1,-k_2;\pi-k',\pi-k'+q')}\nonumber\\
&\times\bigl[G_{2\perp(k_1-k',k_1,k_1-q')}G_{\|(k_2-k',k_2-q',k_2)}G_{1\perp(q-q',k_1-q',k_2-q')}\nonumber\\
&\qquad
+G_{\|(k_1-k',k_1,k_1-q')}G_{2\perp(k_2-k',k_2-q',k_2)}G_{1\perp(q-q',k_1-q',k_2-q')}\bigl],
\\
&\Xi^{(2)}_{2\perp (q,k_1,k_2,q',k')}=J_{2(q+k';k_1,k_2;k',k'-q')}\nonumber\\
&\qquad
\times\bigl[G_{2\perp(q+q',k_1,k_2)}G_{2\perp(q-k_2+k',k'-q',k')}G_{2\perp(q-k_1+k',k',k'-q')}\nonumber\\
&\qquad
+G_{2\perp(q+q',k_1,k_2)}G_{\|(q-k_2+k',k'-q',k')}G_{\|(q-k_1+k',k',k'-q')}\nonumber\\
&\qquad
+G_{\|(q+q',k_1,k_2)}G_{1\perp(q-k_2+k',k'-q',k')}G_{1\perp(q-k_1+k',k',k'-q')}\bigl]\nonumber\\
&\qquad
+J_{2(-k';-k_1,-k_2;\pi-k',\pi-k'+q')}\nonumber\\
&\qquad
\times\bigl[G_{2\perp(k_1-k',k_1,k_1-q')}G_{2\perp(k_2-k',k_2-q',k_2)}G_{2\perp(q-q',k_1-q',k_2-q')}\nonumber\\
&\qquad
+G_{\|(k_1-k',k_1,k_1-q')}G_{\|(k_2-k',k_2-q',k_2)}G_{2\perp(q-q',k_1-q',k_2-q')}\nonumber\\
&\qquad
+G_{1\perp(k_1-k',k_1,k_1-q')}G_{1\perp(k_2-k',k_2-q',k_2)}G_{\|(q-q',k_1-q',k_2-q')}\bigl],
\\
&\Xi^{(2)}_{\| (q,k_1,k_2,q',k')}=J_{2(q+k';k_1,k_2;k',k'-q')}\nonumber\\
&\qquad
\times\bigl[G_{\|(q+q',k_1,k_2)}G_{2\perp(q-k_2+k',k'-q',k')}G_{2\perp(q-k_1+k',k',k'-q')}\nonumber\\
&\qquad
+G_{\|(q+q',k_1,k_2)}G_{\|(q-k_2+k',k'-q',k')}G_{\|(q-k_1+k',k',k'-q')}\nonumber\\
&\qquad
+G_{2\perp(q+q',k_1,k_2)}G_{1\perp(q-k_2+k',k'-q',k')}G_{1\perp(q-k_1+k',k',k'-q')}\bigl]\nonumber\\
&\qquad
+J_{2(-k';-k_1,-k_2;\pi-k',\pi-k'+q')}\nonumber\\
&\qquad
\times\bigl[G_{2\perp(k_1-k',k_1,k_1-q')}G_{2\perp(k_2-k',k_2-q',k_2)}G_{\|(q-q',k_1-q',k_2-q')}\nonumber\\
&\qquad
+G_{\|(k_1-k',k_1,k_1-q')}G_{\|(k_2-k',k_2-q',k_2)}G_{\|(q-q',k_1-q',k_2-q')}\nonumber\\
&\qquad
+G_{1\perp(k_1-k',k_1,k_1-q')}G_{1\perp(k_2-k',k_2-q',k_2)}G_{2\perp(q-q',k_1-q',k_2-q')}\bigl],
\end{align}
\end{subequations}
where $G^2_{\Sigma_n(q,k,k')}$ is given by 
\begin{align}
G^2_{\Sigma_n(q,k,k')}=\sum_{\nu=1\perp,2\perp,\|}G^2_{\nu}(q,k,k')
\end{align}
and  $J_{0(q,k,k')}$, $J_{1(q,k,k')}$ are given  as follows. 
For $|Y^\mathrm{P}_{q,k,k',0}(l)|<E$,
\begin{subequations}
\begin{align}
J_{0(q,k,k')}&
=
2E\ln{\bigl[\frac{4E+Y^\mathrm{P}_{q,k,k',0}(l)}
  {4E-Y^\mathrm{P}_{q,k,k',0}(l)}\bigr]},\\
J_{1(q,k,k')}&=\dfrac{16E^2}{16E^2-(Y^\mathrm{P}_{q,k,k',0}(l))^2}.
\end{align}
\end{subequations}
For $|Y^\mathrm{P}_{q,k,k'}(l)|>E$,
\begin{subequations}
\begin{align}
J_{0(q,k,k')}&=2E\ln{\bigl[\frac{4E+|Y^\mathrm{P}_{q,k,k',0}(l)|}{4E+|Y^\mathrm{P}_{q,k,k',0}(l)|}\bigr]}\mathrm{sgn}(Y^\mathrm{P}_{q,k,k',0}(l)),\\
J_{1(q,k,k')}&=\dfrac{2E}{4E+|Y^\mathrm{P}_{q,k,k',0}(l)|}+\dfrac{2E}{2E+|Y^\mathrm{P}_{q,k,k',0}(l)|}.
\end{align}
\end{subequations}
The quantity $J_{2(q+k'';k_1,k_2;k',k'')}$ is also given by
\begin{align}
J_{2(q+k'';k_1,k_2;k',k'')}=\frac{1}{2}\bigl[J_{1(q+k''-k_1,k',k'')}+J_{1(q+k''-k_2,k',k'')}\bigr].
\end{align}
In the present 
 calculations carried out at the two-loop level, 
we treated  nesting deviations as the dominant effect. 
The  influence of magnetic field 
 at that level remains to be examined. 
This will be the subject of a separate publication.


\begin{thebibliography}{99} 
\bibitem{Gruner} 
G. Gr\"uner: 
Density Waves in Solids
(Addison Wesley, Massachusetts and Tokyo, 2007).
\bibitem{Kuroki} 
K. Kuroki: 
J. Phys. Soc. Jpn. \textbf{75} (2006)  051013.
\bibitem{Lee} 
I. J. Lee, M. J. Naughton, G. M. Danner, and P. M. Chaikin:
Phys. Rev. Lett. \textbf{78} (1997) 3555. 
\bibitem{Zhang} 
W. Zhang and C.A.R. S\'a de Melo: 
Adv. Phys. \textbf{56} (2007) 545.
\bibitem{Shinagawa1} 
J. Shinagawa, Y. Kurosaki, F. Zhang, C. Parker, S. E. Brown, 
D. J\'erome, K. Bechgaard, and J. B. Christensen: 
Phys. Rev. Lett. \textbf{98} (2007) 147002. 
\bibitem{Yonezawa1} 
S. Yonezawa, S. Kusaba, Y. Maeno, P. Auban-Senzier, 
C. Pasquier, K. Bechgaard, and D. J\'erome: 
Phys. Rev. Lett. \textbf{100} (2008) 117002.
\bibitem{Wzietek93} 
P. Wzietek, F. Creuzet, C. Bourbonnais, D. J\'erome, 
K. Bechgaard, and P. Batail: 
J. Phys. I {\bf 3} (1993) 171.
\bibitem{Schimahara2000JPSJ} 
H. Shimahara: 
J. Phys. Soc. Jpn. \textbf{69} (2000) 1966.
\bibitem{Schimahara2000PRB} 
H. Shimahara: 
Phys. Rev. B \textbf{62} (2000) 3524.
\bibitem{Belmechri1} 
N. Belmechri, G. Abramovici, M. Heritier, 
S. Haddad, and S. Charfi-Kaddour: 
Eur. Phys. Lett. \textbf{80} (2007) 37004.
\bibitem{Belmechri2} 
N. Belmechri, G. Abramovici, and M. Heritier: 
Eur. Phys. Lett. \textbf{82} (2008) 47009. 
\bibitem{Aizawa1} 
H. Aizawa, K. Kuroki, and Y. Tanaka: 
Phys. Rev. B \textbf{77} (2008) 144513.  
\bibitem{Aizawa2} 
H. Aizawa, K. Kuroki, T. Yokoyama, and Y. Tanaka: 
Phys. Rev. Lett. \textbf{102} (2009) 016403.
\bibitem{Scalapino1987} 
D.J. Scalapino, E. Loh, Jr., and J.E. Hirsch:
Phys. Rev. B \textbf{35} (1987) 6694.
\bibitem{Schimahara1989} 
H. Shimahara: 
J. Phys. Soc. Jpn. \textbf{58} (1989) 1735.
\bibitem{Tanaka2004} 
Y. Tanaka and K. Kuroki:
 Phys. Rev. B \textbf{70} (2004) 060502.
\bibitem{Solyom} 
J. S\'olyom: 
Adv. Phys. \textbf{28} (1979) 201.
\bibitem{Bourbonnais1} 
C. Bourbonnais, B. Guay, and R. Wortis: 
in Theoretical Methods for Strongly Correlated Electrons, 
edited by D. Se\'ne\'chal, A.M. Tremblay, and C. Bourbonnais 
(Springer, New York, 2003) p. 77.
\bibitem{Penc} 
K. Penc and J. S\'olyom: 
Phys. Rev. B \textbf{47} (1993) 6273.
\bibitem{Montambaux} 
G. Montambaux, M. H\'eritier, and P. Lederer: 
Phys. Rev. B \textbf{33} (1986) 7777.
\bibitem{D-B} 
R. Duprat and C. Bourbonnais: 
 Eur. Phys. J. B. \textbf{21} (2001)  219.
\bibitem{Fuseya1} 
Y. Fuseya and Y. Suzumura: 
J. Phys. Soc. Jpn. \textbf{74} (2005) 1263.
\bibitem{Nickel1} 
J. C. Nickel, R. Duprat, C. Bourbonnais, and N. Dupuis: 
Phys. Rev. Lett. \textbf{95}  (2005) 247001.
\bibitem{Tsuchiizu1} 
M. Tsuchiizu: Phys. Rev. B. \textbf{74} (2006) 155109.
\bibitem{Abramovici1} 
G. Abramovici, J. C. Nickel, and M. H\'eritier:
Phys. Rev. B. \textbf{72}  (2005)  045120.
\bibitem{SuzumuraPRB} 
Y. Suzumura, M. Tsuchiizu, and G. Gr\"uner: 
Phys. Rev. B \textbf{57} (1998) R15040.
\bibitem{TsuchiizuPRB} 
M. Tsuchiizu and Y. Suzumura: 
Phys. Rev. B \textbf{59} (1999) 12326.
\bibitem{Emery} 
V.J. Emery, R. Bruinsma, and S. Barisi\'c: 
Phys. Rev. Lett. \textbf{48} (1982) 1039.
\bibitem{Tsuchiizu2005} 
M. Tsuchiizu and Y. Suzumura:
 Phys. Rev. B \textbf{72} (2005) 075121.
\bibitem{Fuseya2005} 
Y. Fuseya, M. Tsuchiizu, Y. Suzumura, and C. Bourbonnais: 
J. Phys. Soc. Jpn. \textbf{74} (2005) 3159.
\bibitem{Fuseya2007} 
Y. Fuseya, M. Tsuchiizu, Y. Suzumura, and C. Bourbonnais: 
J. Phys. Soc. Jpn. \textbf{76} (2007) 014709.
\bibitem{Fabrizio1} 
M. Fabrizio: 
Phys. Rev. B. \textbf{48} (1993) 15838.
\end{thebibliography}
\end{document}